\documentclass[aip,jcp,groupedaddress,reprint]{revtex4-1}
\usepackage{amsmath,amssymb,bm}
\usepackage{graphicx,tabularx,dcolumn}
\usepackage{xcolor}
\usepackage{amsmath}
\usepackage{amsfonts}
\usepackage{amssymb}
\usepackage{graphicx}

\graphicspath{{./figures/}}
\newcolumntype{d}[1]{D{.}{.}{#1}}
\colorlet{DarkGreen}{green!70!black}
\begin{document}

\title{Improving the accuracy and efficiency of time-resolved electronic
spectra calculations: Cellular dephasing representation with a prefactor}
\author{Eduardo Zambrano}
\author{Miroslav \v{S}ulc}
\author{Ji\v{r}\'{\i} Van\'{\i}\v{c}ek}
\email{jiri.vanicek@epfl.ch}
\affiliation{Laboratory of Theoretical Physical Chemistry, Institut des Sciences et Ing%
\'{e}nierie Chimiques, \'{E}cole Polytechnique F\'{e}d\'{e}rale de Lausanne
(EPFL), CH-1015 Lausanne, Switzerland}
\date{\today}

\begin{abstract}
\noindent Time-resolved electronic spectra can be obtained as the Fourier
transform of a special type of time correlation function known as fidelity
amplitude, which, in turn, can be evaluated approximately and efficiently
with the dephasing representation. Here we improve both the accuracy of this
approximation---with an amplitude correction derived from the phase-space
propagator---and its efficiency---with an improved cellular scheme employing
inverse Weierstrass transform and optimal scaling of the cell size. We
demonstrate the advantages of the new methodology by computing dispersed
time-resolved stimulated emission spectra in the harmonic potential,
pyrazine, and the NCO molecule. In contrast, we show that in strongly
chaotic systems such as the quartic oscillator the original dephasing
representation is more appropriate than either the cellular or
prefactor-corrected methods.
\end{abstract}

\maketitle

\section{Introduction}

Ultrafast spectroscopy with a time resolution as high as $10^{-15}\,$s is
essential for understanding many quantum dynamical processes in chemical
physics.\cite{Bisgaard_Clarkin:2009,*Bressler_Milne:2009,*Carbone_Kwon:2009}
Although short time scales should simplify theoretical studies by requiring
shorter simulations, solving the time-dependent Schr\"{o}dinger equation
(TDSE) is challenging even for short times due to the exponential scaling
with the number of degrees of freedom. An attractive approach offering a
compromise between accuracy and computational efficiency is provided by the
semiclassical initial value representation methods,\cite%
{Miller:1970,Heller:1981,Herman_Kluk:1984,Miller:2001,Herman:1994,*Thoss_Wang:2004,*Kay:2005,*Ceotto_Zhuang:2013}
which benefit from the ultrafast character of the dynamics not only because
of lower computational cost, but also because their accuracy deteriorates
with increasing time.

The so-called dephasing representation\cite{Vanicek:2004,*Vanicek:2006}
(DR), is an efficient initial-value-type semiclassical approximation
particularly fitted for calculations of time-resolved electronic spectra.%
\cite{Wehrle_Sulc:2011,Sulc_Vanicek:2012} The DR improves on a previous
method\cite{Vanicek_Heller:2003} inspired by the semiclassical perturbation
theory of Miller and coworkers.\cite{Miller_Smith:1978,*Hubbard_Miller:1983}
In electronic spectroscopy, the DR and closely related approximations are
known as Mukamel's phase averaging method\cite{Mukamel:1982,*book_Mukamel}
or Wigner-averaged classical limit, and were used by various authors.\cite%
{Shemetulskis_Loring:1992,Rost:1995,Wang_Sun:1998,Li_Fang:1996,*Egorov_Rabani:1998,*Egorov_Rabani:1999,Shi_Geva:2005}
Shi and Geva\cite{Shi_Geva:2005} derived this approximation without invoking
the semiclassical propagator---by linearizing\cite%
{Poulsen_Nyman:2003,Bonella_Coker:2005} the path integral quantum
propagator. Although the original formulation of the DR pertains to a single
pair of potential energy surfaces, the generalization to multiple surfaces,
and hence to nonadiabatic dynamics, exists.\cite%
{Zimmermann_Vanicek:2012,*Zimmermann_Vanicek:2012a} The DR has many other
applications; the method successfully described, e.g., the local density of
states and the transition from the Fermi-Golden-Rule to the Lyapunov regime
of fidelity decay.\cite%
{Wang_Casati:2005,*Ares_Wisniacki:2009,*Wisniacki_Ares:2010,*Garcia-Mata_Wisniacki:2011}

Yet the most attractive feature of the DR is its efficiency: Motivated by
numerical comparisons with other semiclassical methods,\cite%
{Wehrle_Sulc:2011} it has been recently proved analytically\cite%
{Mollica_Vanicek:2011} that the number of trajectories required for
convergence of the DR is independent of the system's dimensionality,
Hamiltonian, or total evolution time. The efficiency was further increased
in the cellular version of the DR,\cite{Sulc_Vanicek:2012} which was
inspired by Heller's cellular dynamics\cite{Heller:1991} and which can
significantly reduce the required number of trajectories. The original
implementation of the \emph{cellular DR} (CDR), however, does not converge
to the DR in the limit of infinite number of trajectories.

Unlike its efficiency, the accuracy of the DR is not always sufficient. The
DR is exact in displaced harmonic oscillators\cite%
{Mukamel:1982,*book_Mukamel} and often accurate in chaotic systems,\cite%
{Vanicek:2004,*Vanicek:2006} but it breaks down in as simple systems as
harmonic oscillators with different force constants. This breakdown can be
partially remedied by augmenting the DR with a~prefactor,\cite%
{Zambrano_Almeida:2011} which, however, leads to a~much higher computational
cost per trajectory and also typically requires more trajectories to achieve
convergence.

The first goal of the present paper is to describe a general numerical
implementation of the prefactor correction and apply it to the calculation
of time-resolved electronic spectra. As the numerical evaluation of the CDR
requires, incidentally, the same ingredients as the prefactor correction,
the second goal is to combine the advantages of the cellular approach and
prefactor correction into a single formula, and show that the resulting
method, \emph{cellular DR with prefactor} (CDRP), is able to increase both
the efficiency and accuracy of the DR. Our third goal is presenting a major
improvement of the cellularization process by employing the inverse
Weierstrass transform of the initial state as the optimal sampling weight
instead of the widely used Wigner or Husimi functions, and by correlating
the size of the cells with their number and the number of degrees of
freedom, which guarantees the convergence of the CDR to the original DR in
the limit of infinite number of trajectories.

The remainder of the paper is organized as follows: The correlation function
approach and the DR approximation for evaluating time-resolved stimulated
emission spectra is reviewed in Section \ref{sec:Theory}; in particular, the
DR, its prefactor correction, and its cellular version are deduced. After
explaining how the new cellular approach provides optimal choices of the
sampling weight and width of Gaussian cells, we derive the CDRP, i.e., a
method combining the prefactor correction and cellularization into a single
framework. Section \ref{sec:Examples} contains several analytical and
numerical results testing the theory developed in Section \ref{sec:Theory},
while Section \ref{sec:Final_remarks} provides conclusions. 

\section{Theory\label{sec:Theory}}

\subsection{Time-resolved stimulated emission: spectrum, time correlation
function, and dephasing representation.\label{subsec:DR}}

To be specific, we will present the methodology for time-resolved stimulated
emission (TRSE); modification to other ultrafast processes is
straightforward. Within the electric dipole approximation, time-dependent
perturbation theory, and ultrashort pulse approximation, the dispersed\cite%
{Pollard_Lee:1990,Stock_Domcke:1992} TRSE spectrum can be computed as a
Fourier transform of the following correlation function:\cite%
{Pollard_Lee:1990,Stock_Domcke:1992,Wehrle_Sulc:2011,Sulc_Vanicek:2012}

\begin{equation}
\begin{split}
C_{\text{TRSE}}(t,\tau )=& E_{\text{pu}}^{2}E_{\text{pr}}\text{Tr}\,\left[ 
\hat{\rho}_{g}(T){\hat{\mu}}_{ge}\hat{U}_{e}(\tau +t)^{-1}\right. \\
& \left. \times {\hat{\mu}}_{eg}\hat{U}_{g}(t){\hat{\mu}}_{ge}\hat{U}%
_{e}(\tau ){\hat{\mu}}_{eg}\right] .
\end{split}
\label{eq:C TRSE}
\end{equation}%
Here $E_{\text{pu}}$ and $E_{\text{pr}}$ denote the amplitudes of the pump
and probe laser pulses, $\hat{\rho}_{g}(T)$ represents the nuclear density
operator in the electronic ground state at temperature $T$, $\hat{\mu}_{ij}$
is the transition dipole moment operator coupling electronic states $i$ and $%
j$, $\tau $ stands for the time delay between the pump and probe pulses, and 
$t$ is the time elapsed after the probe pulse. Finally, $\hat{U}_{j}$
denotes the nuclear quantum evolution operator 
\begin{equation}
\hat{U}_{j}=\exp (-i\hat{H}_{j}t/\hbar )\quad \quad (j=g,\,e),
\label{eq:U for e and g}
\end{equation}%
with Hamiltonian $\hat{H}_{j}=\hat{T}+\hat{V}_{j}$ where $\hat{T}$ is the
nuclear kinetic energy and $\hat{V}_{j}$ is the $j$th potential energy
surface (PES). In all expressions, the hat denotes operators in the Hilbert
space of nuclei.

Within the Franck-Condon approximation and zero-temperature limit,
correlation function (\ref{eq:C TRSE}) reduces to 
\begin{equation}
C_{\text{TRSE}}(t,\tau )=E_{\text{pu}}^{2}E_{\text{pr}}|\mu
_{eg}|^{4}f(t,\tau ),  \label{eq:C_TRSE_approx}
\end{equation}%
where 
\begin{align}
f(t,\tau ):=& \langle \psi _{e}(t,\tau )|\psi _{g}(t,\tau )\rangle ,
\label{eq:f(t,tau)} \\
|\psi _{j}(t,\tau )\rangle :=& \hat{U}_{j}(t)\hat{U}_{e}(\tau )|\Psi _{\text{%
init}}\rangle ,
\end{align}%
is a specific time correlation function and the initial state $|\Psi _{\text{%
init}}\rangle $ is typically the vibrational ground state of the ground PES.
The TRSE spectrum, given by\cite{Pollard_Lee:1990} 
\begin{equation}
\sigma _{\text{TRSE}}(\omega ,\tau )\propto \omega E_{\text{pu}}^{2}E_{\text{%
pr}}|\mu _{eg}|^{4}\sigma (\omega ,\tau ),  \label{eq:sigma_TRSE}
\end{equation}%
is proportional to the so-called wave packet spectrum $\sigma $ obtained\cite%
{book_Tannor} as 
\begin{equation}
\sigma (\omega ,\tau )=\text{Re}\,\int_{0}^{\infty }\!\!dt\,f(t,\tau
)e^{i\omega t}.  \label{eq:sigma_wp}
\end{equation}%
Correlation function (\ref{eq:f(t,tau)}) for the stimulated emission is a
particular example of a more general concept of \emph{fidelity amplitude},%
\cite{Gorin_Prosen:2006,*Jacquod_Petitjean:2009} defined as 
\begin{equation}
f(t)=\langle \Psi _{\text{init}}|\hat{U}_{1}(t,0)^{-1}\hat{U}_{2}(t,0)|\Psi
_{\text{init}}\rangle ,  \label{eq:fid_amplitude}
\end{equation}%
where $U_{J}(t_{2},t_{1})$, $J=1,\,2$, is the time evolution operator for a
time-dependent Hamiltonian $\hat{H}_{J}(t)$: 
\begin{equation}
\hat{U}_{J}(t_{2},t_{1})=\mathcal{T}\,\exp \left[ -\frac{i}{\hbar }%
\int_{t_{1}}^{t_{2}}dt^{\prime }\hat{H}_{J}(t^{\prime })\right] \quad \quad
(J=1,2),  \label{eq:U_J}
\end{equation}%
where $\mathcal{T}$ denotes the time-ordering operator.

Correlation function (\ref{eq:f(t,tau)}) for TRSE is obtained from the
general fidelity amplitude~(\ref{eq:fid_amplitude}) by substituting the
following time-dependent Hamiltonians $\hat{H}_{J}(t)$ into Eq.~(\ref{eq:U_J}%
):%
\begin{equation}
\begin{split}
\hat{H}_{1}(t^{\prime}) & \equiv\,\,\,\,\,\hat{H}_{e}\quad\text{ for }0\leq
t^{\prime}\leq\tau+t, \\
\hat{H}_{2}(t^{\prime}) & \equiv%
\begin{cases}
\hat{H}_{e} & \text{ for }0\leq t^{\prime}\leq\tau, \\ 
\hat{H}_{g} & \text{ for }\tau\leq t^{\prime}\leq\tau+t.%
\end{cases}%
\end{split}
\label{eq:H1}
\end{equation}
Note that $\hat{H}_{2}(t^{\prime})\equiv\hat{H}_{g}$ if $\tau=0$.

Besides electronic spectroscopy applications,\cite%
{Li_Fang:1996,*Egorov_Rabani:1998,*Egorov_Rabani:1999,Shi_Geva:2005,Rost:1995,Shemetulskis_Loring:1992}
correlation function (\ref{eq:fid_amplitude}) proved useful, e.g., in NMR
spin echo experiments\cite{Pastawski_Levstein:2000} and in the theories of
quantum computation, decoherence,\cite%
{Cucchietti_Dalvit:2003,*Gorin_Prosen:2004} and inelastic neutron scattering.%
\cite{Petitjean_Bevilaqua:2007} Fidelity amplitude was also used as a
measure of the dynamical importance of diabatic, nonadiabatic, or spin-orbit
couplings,\cite%
{Zimmermann_Vanicek:2010,Zimmermann_Vanicek:2012,*Zimmermann_Vanicek:2012a}
and of the accuracy of quantum molecular dynamics on an approximate PES.\cite%
{Li_Mollica:2009,Zimmermann_Ruppen:2010}

In practical calculations, correlation function (\ref{eq:fid_amplitude})
must usually be approximated, and DR provides an efficient semiclassical
approximation.\cite%
{Vanicek:2004,*Vanicek:2006,Li_Fang:1996,*Egorov_Rabani:1998,*Egorov_Rabani:1999,Shi_Geva:2005,Rost:1995,Shemetulskis_Loring:1992}
If we denote by $x^{t}:=(q^{t},p^{t})$ the phase-space coordinates at time $%
t $ of a point along a classical trajectory of the \emph{average}\cite%
{Wehrle_Sulc:2011,Mukamel:1982,*book_Mukamel,Zambrano_Almeida:2011}
Hamiltonian $H:=(H_{1}+H_{2})/2$, the DR of fidelity amplitude (\ref%
{eq:fid_amplitude}) can be written as 
\begin{equation}
f_{\text{DR}}(t)=h^{-D}\int dx^{0}\,\rho_{W}(x^{0})e^{iS_{\text{DR}%
}(x^{0},t)/\hbar },  \label{eq:DR formula integral}
\end{equation}%
with 
\begin{equation}
\rho_{W}(q,p)\equiv \int ds\left\langle q-s/2\right\vert \hat{\rho}_{\text{%
init}}\left\vert q+s/2\right\rangle e^{is^{\mathsf{T}}\cdot p/\hbar }.
\label{eq:Wigner transform}
\end{equation}%
Here $D$ is the number of degrees of freedom, $\rho_{W}$ denotes the Wigner
transform of the initial density operator $\hat{\rho}_{\text{init}}=|\Psi _{%
\text{init}}\rangle \langle \Psi _{\text{init}}|$, and $S_{\text{DR}%
}(x^{0},t)$ is the action due to the difference $\Delta H:=H_{2}-H_{1}$
along trajectory $x^{t}$: 
\begin{equation}
S_{\text{DR}}({x}^{0},t)=-\int_{0}^{t}dt^{\prime }\,\Delta H({x}^{t^{\prime
}},t^{\prime }).  \label{eq:DR phase}
\end{equation}%
For TRSE, $\Delta H$ is given by 
\begin{equation}
\Delta H\equiv 
\begin{cases}
0 & \text{for }0\leq t^{\prime }\leq \tau , \\ 
V_{g}-V_{e} & \text{for }\tau \leq t^{\prime }\leq \tau +t.%
\end{cases}%
\end{equation}%
Denoting the phase-space average of a quantity $A(x)$ with respect to a
weight function $w(x)$ by 
\begin{equation}
\left\langle A(x)\right\rangle _{w(x)}:=\frac{\int d{x}A(x)w(x)}{\int d{x}%
\,w(x)},  \label{eq:<  >}
\end{equation}%
time correlation function (\ref{eq:DR formula integral}) can be written in a
compact way as 
\begin{equation}
f_{\text{DR}}(t)=\left\langle e^{iS_{\text{DR}}(x^{0},t)/\hbar
}\right\rangle _{\rho_{W}(x^{0})}.  \label{eq:DR formula}
\end{equation}%
Formula (\ref{eq:DR formula}) can be evaluated efficiently by Monte Carlo
integration. Indeed, because the convergence of the DR is independent of
dimensionality, the DR is in many-dimensional systems much more efficient
than other quantum or classical algorithms for computing the fidelity
amplitude.\cite{Mollica_Vanicek:2011} The accuracy of the DR typically
improves with decreasing $\Delta H$ and increasing complexity of
Hamiltonians $H_{1}$ and $H_{2}$. While the DR is exact in displaced
harmonic oscillators with arbitrary displacement, this perturbative
approximation breaks down in some singular cases, such as when Hamiltonians $%
H_{1}$ and $H_{2}$ represent harmonic oscillators with significantly
different force constants.\cite{Mukamel:1982}

\subsection{Prefactor correction\label{subsec:DRP}}

The above-mentioned breakdown of the DR can be partially corrected by
including a prefactor in the DR formula~(\ref{eq:DR formula}).\cite%
{Zambrano_Almeida:2011} We now briefly derive this improved version of the
DR.

\begin{figure}[htbp]
\includegraphics[width=0.9\columnwidth]{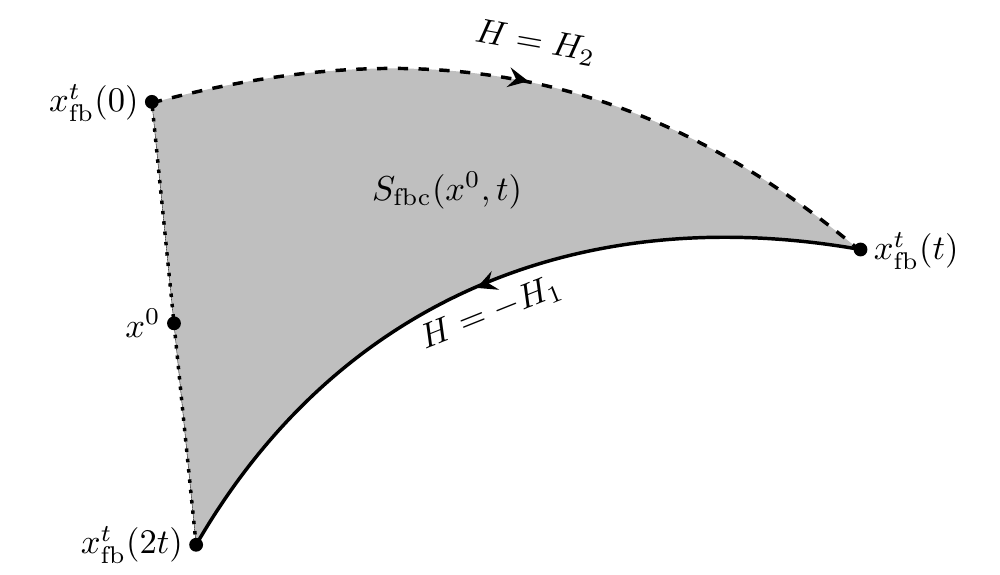} 
\caption{Sketch of semiclassical evaluation of fidelity amplitude in phase space. Given a phase-space point ${x}^{0}$, the path ${x}_{\text{fb}}^{t'}$ is
determined by two requirements: (\emph{i}) for $0\leq t' \leq t$ it is
driven by $H_{2}$ (dashed path), while for $t\leq t' \leq 2t$ it is
driven by $-H_{1}$ (continuous path); and (\emph{ii}) ${x}^{0}=(x_{\text{fb}}^{0}+x_{\text{fb}}^{t})/2$.
Geometrical part of the phase $S_{\text{fbc}}({x}^{0},t)$ is the shadowed area and the dotted line is the chord between $x^0_{\text{fb}}$ and $x^t_{\text{fb}}$.}
\label{fig:DRP Figure}
\end{figure}

Fidelity amplitude (\ref{eq:fid_amplitude}) can be expressed as the
expectation value of the \emph{echo operator}\cite%
{Gorin_Prosen:2006,*Jacquod_Petitjean:2009} $\hat{\mathcal{E}}(t):=\hat{U}%
_{1}(t,0)^{-1}\hat{U}_{2}(t,0)$: 
\begin{equation}
f(t)=\text{Tr}\,\left[ \hat{\rho}\,\hat{\mathcal{E}}(t)\right] =\left\langle 
\mathcal{E}_{W}(x^{0},t)\right\rangle _{\rho_{W}(x^{0})},  \label{eq:f trace}
\end{equation}%
where $\mathcal{E}_{W}(x^{0},t)$ is the Wigner transform of the echo
operator. Note that $\hat{\mathcal{E}}(t)$ itself can be interpreted as a
single \textquotedblleft forward-backward\textquotedblright\ evolution
operator describing propagation driven by $H_{2}$ for time $t$ followed by a
propagation driven by $-H_{1}$ from time $t$ to $2t$. The path labeled by $%
x_{\text{fb}}^{t}(t^{\prime })$ in Fig.~\ref{fig:DRP Figure} is a classical
analog of such a forward-backward propagation.

A semiclassical approximation to the Wigner transform $\mathcal{E}_{W}({x}%
^{0},t)$ consists in replacing it by a single phase-space semiclassical
propagator,\cite{Berry:1989,Almeida:1998} 
\begin{equation}
\mathcal{E}_{\text{SC}}(x^{0},t)=\left\vert \det \left( I+\frac{J}{2}\cdot 
\frac{\partial ^{2}S_{\text{fbc}}}{\partial (x^{0})^{2}}\right) \right\vert
^{\frac{1}{2}}e^{iS_{\text{fbc}}(x^{0},t)/\hbar },  \label{eq:Echo_symbol}
\end{equation}%
with the constraint $x^{0}=[x_{\text{fb}}^{t}(2t)+x_{\text{fb}}^{t}(0)]/2$.
Here $I$ is the identity matrix in $2D$ dimensions and $J$ is the standard
symplectic matrix in $2D$ dimensions, 
\begin{equation}
J=%
\begin{pmatrix}
0_{D} & I_{D} \\ 
-I_{D} & 0_{D}%
\end{pmatrix}%
,
\end{equation}%
where the subscripts specify the dimensionality of each square block. More
details about this semiclassical phase-space propagator are presented in
Appendix \ref{ap:PS propagator}. In Eq.~(\ref{eq:Echo_symbol}), phase $S_{%
\text{fbc}}(x^{0},t)$ is the so-called \emph{center-action} of the path $x_{%
\text{fb}}^{t}(t^{\prime })$ at time $t$; explicitly, this function is
defined as 
\begin{equation}
S_{\text{fbc}}(x^{0},t):=\oint p^{\mathsf{T}}\cdot
dq-\int_{0}^{2t}\!dt^{\prime }H(x_{\text{fb}}^{t}(t^{\prime}), t^{\prime}),
\label{eq:S_fb}
\end{equation}%
where the closed integral is evaluated along the path consisting of $x_{%
\text{fb}}^{t}(t^{\prime })$ and of the straight line connecting $x_{\text{fb%
}}^{t}(2t)$ and $x_{\text{fb}}^{t}(0)$, as shown in Fig.~\ref{fig:DRP Figure}%
, and 
\begin{equation}
H(x_{\mathrm{fb}}^{t}(t^{\prime }),t^{\prime })\equiv 
\begin{cases}
H_{2}(x_{\mathrm{fb}}^{t}(t^{\prime }),t^{\prime }) & \text{for }0\leq
t^{\prime }\leq t, \\ 
-H_{1}(x_{\mathrm{fb}}^{t}(t^{\prime }),2t-t^{\prime }) & \text{for }t\leq
t^{\prime }\leq 2t.%
\end{cases}%
\end{equation}%
Center-action (\ref{eq:S_fb}) appears naturally in the Weyl representation
of quantum mechanics.\cite{Almeida:1998} As mentioned in Appendix \ref{ap:PS
propagator}, the center-action is a function of the center $x^{0}$ and, in
general, is multivalued: a given center $x^{0}$ may be the midpoint between
the initial and final points for two or more paths (see, e.g., Fig.~\ref%
{fig:LE-areas U(x)} in Appendix \ref{ap:PS propagator}). Nevertheless, as
shown in Appendix~\ref{ap:PS propagator}, for our purposes, we can assume
that $S_{\text{fbc}}(x^{0},t)$ has only a single branch.

Approximating the center-action in the semiclassical echo operator~(\ref%
{eq:Echo_symbol}) by the DR action, ${S}_{\text{fbc}}(x^{0},t)\simeq S_{%
\text{DR}}(x^{0},t)$, which is valid up to the first order in perturbation
theory,\cite{Bohigas_Giannoni:1995,Zambrano_Almeida:2011} yields an improved
approximation for fidelity amplitude given by $f(t)\approx f_{\text{DRP}}(t)$%
, where 
\begin{equation}
f_{\text{DRP}}(t)=\left\langle A_{\text{DRP}}(x^{0},t)e^{iS_{\text{DR}%
}(x^{0},t)/\hbar}\right\rangle _{\rho_{W}(x^{0})},  \label{eq:DRP formula}
\end{equation}
with 
\begin{align}
A_{\text{DRP}}(x^{0},t) & :=\left\vert \det\left( I+J\cdot
B_{x^{0}}^{t}\right) \right\vert ^{1/2},  \label{eq:A_DRP} \\
B_{{x}^{0}}^{t} & \equiv B(x^{0},t):=\frac{1}{2}\frac{\partial ^{2}S_{\text{%
DR}}(x^{0},t)}{\partial(x^{0})^{2}}.  \label{eq:dS_expansion_B}
\end{align}

We will refer to expression (\ref{eq:DRP formula}) as the \emph{DR with
prefactor} or DRP: it corresponds to including a prefactor to the
contribution of each trajectory in the DR formula~(\ref{eq:DR formula}). The
DRP is free of caustics because the prefactor~(\ref{eq:A_DRP}) cannot
diverge. However, the prefactor is the most expensive part of the DRP
evaluation because it depends on the Hessian of the DR phase $S_{\text{DR}%
}(x^{0},t)$ with respect to the initial conditions; in Appendix~\ref%
{ap:Derivatives} we show how to compute this Hessian from the derivatives of
the stability matrix of the classical trajectory. Finally note that
switching the PESs in the definition~(\ref{eq:fid_amplitude}) of fidelity
amplitude is equivalent to taking the complex conjugate of this equation.
DRP preserves this property because of the identity $\det (I+J\cdot
B_{x^{0}}^{t})=\det (I-J\cdot B_{x^{0}}^{t})$, proven in Appendix \ref%
{ap:Proof}. 

\subsection{Cellularization\label{subsec:CDR}}

The \emph{cellular dephasing representation} (CDR) was developed in Ref.~%
\onlinecite{Sulc_Vanicek:2012} in order to further accelerate the
convergence of the DR in the spirit of Heller's \emph{cellular dynamics}.%
\cite{Heller:1991} The main idea of the CDR consists in decomposing the
Wigner transform of the initial state into phase-space cells and evaluating
the contribution of an entire cell of nearby trajectories approximately,
using the dynamical information collected along a single, central
trajectory. Here we describe a simpler and more rigorous cellularization
process than that used in the original CDR (Ref.~%
\onlinecite{Sulc_Vanicek:2012}) and other cellularization\cite%
{Heller:1991,Walton_Manolopoulos:1996} or Filinov filtering\cite%
{Filinov:1986,Makri_Miller:1987,Wang:2001} schemes. In particular, the new
methodology provides both a natural criterion for cell size [see Eq.~(\ref%
{eq:lambda_scaling})] and a~natural sampling weight for the cell centers
[given by inverse Weierstrass transform (\ref{eq:weierstrass trans})]. Most
importantly, unlike the previous approaches, in the limit of infinite number
of trajectories, the new methodology converges to the original, noncellular
method (in our case, the DR).

In standard cellularization or Filinov filtering procedures,\cite%
{Filinov:1986,Makri_Miller:1987,Heller:1991,Walton_Manolopoulos:1996,Wang:2001,Sulc_Vanicek:2012}
the initial state is covered with phase-space Gaussians as in Fig.~\ref%
{fig:cellularization}(a), the centers of these Gaussians being sampled from
a given distribution (denoted with a black circle), typically a Wigner or
Husimi transform of the initial state, which is independent of the size and
number of cells. Then one decreases the cell size (measured by parameter $%
\lambda $, defined so that each cell has phase-space volume $\lambda
^{2D}h^{D}$) until the approximate treatment of contributions of neighboring
trajectories (typically involving quadratic expansion of the action) becomes
sufficiently accurate. Independently, the number of cells $N$ is increased
until convergence.

\begin{figure*}[htbp]
\includegraphics[width=0.7\hsize]{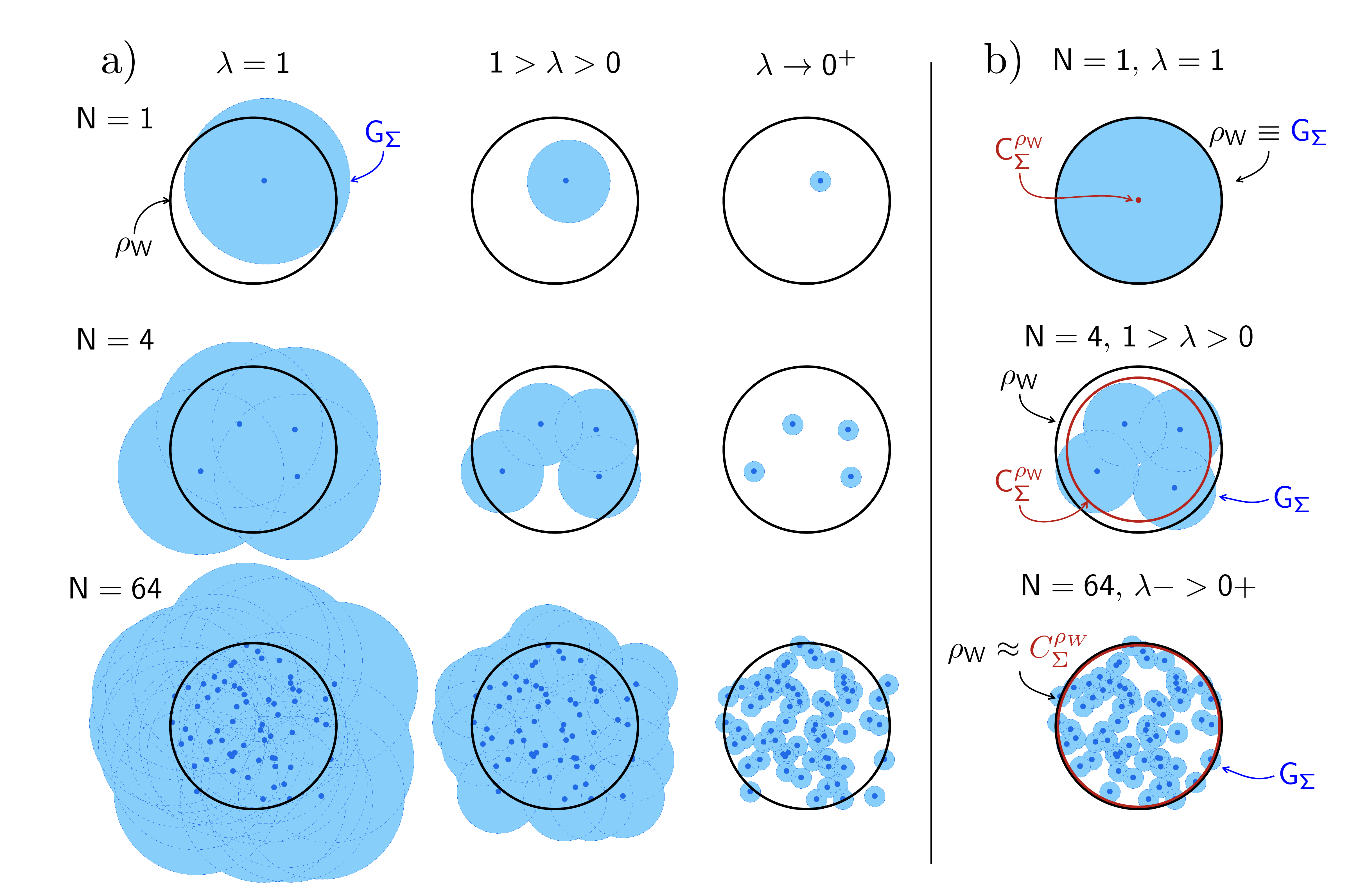}
\caption{Comparison of the standard (a) and new (b) cellularization schemes.
In both panels, black circles represent the initial state, while the
light-blue disks are the Gaussian cells. (a) In the standard procedure, the
number of cells $N$ and their size $\protect\lambda \in (0,1]$ are
independent. The sampling weight for the cell centers, given by the Wigner
function $\protect\rho_{W}$ (black circle), is independent of both $N$ and $%
\protect\lambda $. (b) In the cellularization procedure proposed in the main
text, both the cell size and the sampling weight for their centers are
uniquely determined by $N$. The weight, given by inverse Weierstrass
transform $C_{\Sigma }^{\protect\rho_{W}}$, is denoted with red circles. }
\label{fig:cellularization}
\end{figure*}

There are several problems with this standard approach: First, decreasing
the size of the cell to zero ($\lambda \rightarrow 0$) for a fixed number of
cells $N$ eventually results in the initial state not being fully covered
[see the middle row of Fig.~\ref{fig:cellularization}(a)]. Second, in case
that the quadratic expansion of the action is accurate, taking the limit $%
N\rightarrow \infty $ for a fixed nonzero width $\lambda $ is wasteful since
many cells are overlapping [see the middle column of Fig.~\ref%
{fig:cellularization}(a)]. Third, if the quadratic expansion is inaccurate,
taking the limit $N\rightarrow \infty $ for a fixed width $\lambda $
converges to a result different from the original noncellular method.
Fourth, for Gaussian initial states and $N=1$, the optimal choice of a
single cell is clearly the initial state, but in the standard approach the
width and position of the cell are uncorrelated with the number of cells
[see the top row of Fig.~\ref{fig:cellularization}(a)].

The solution of the first three problems is simple and provided by scaling
the size of the cell with the number of cells and dimensions according to%
\begin{equation}
\lambda =N^{-1/2D},  \label{eq:lambda_scaling}
\end{equation}%
guaranteeing that the phase-space volume of the initial state is equal to
the total volume of all cells [Fig.~\ref{fig:cellularization}(b)]. This
avoids an \textit{ad hoc} choice of the width of the cell, replacing two
limiting processes $\lambda \rightarrow \infty $ and $N\rightarrow \infty $
with a single process $N\rightarrow \infty $, and pictorially corresponds to
going along the diagonal from the top left to the bottom right corner of
Fig.~\ref{fig:cellularization}(a). In the derivation presented below it is
shown that the fourth problem is solved by sampling the centers of the cells
from the inverse Weierstrass instead of the Wigner transform of the initial
state. As we shall see, this inverse Weierstrass transform, represented by
red circles in Fig.~\ref{fig:cellularization}(b) is a natural sampling
weight, which is correlated to the size of the cell. If the initial state is
a Gaussian, for $N=1$, the single cell has uniquely defined size and
position, equal to the size and position of the initial state. In the limit
of infinitely many very small cells, their centers are sampled from the
Wigner transform. All together, $N$ determines both the size of each cell
and the sampling weight for their centers.

To put the above ideas into a precise mathematical form, consider
a~phase-space Gaussian function centered at the origin, 
\begin{equation}
G_{\Sigma }(x):=\hbar ^{D}\sqrt{\det \Sigma }\,\,e^{-x^{\mathsf{T}}\cdot {%
\Sigma }\,\cdot x/2},  \label{eq:Weyl Gaussian}
\end{equation}%
where $\Sigma $ is a~$2D\times 2D$ real, symmetric, positive definite
matrix, whose determinant is inversely proportional to the square of the
phase-space volume occupied by $G_{\Sigma }$, while the prefactor in Eq.~(%
\ref{eq:Weyl Gaussian}) ensures normalization of $G_{\Sigma }$: $h^{-D}\int
\!dx\,G_{\Sigma }(x)=1$. In particular, if $\Sigma _{i,i}=2/\sigma ^{2}$ and 
$\Sigma _{D+i,D+i}=2\sigma ^{2}/\hbar ^{2}$ (for $i=1,\dots ,D$ and $\sigma
>0$), then $G_{\Sigma }(x)$ coincides with the Wigner transform of a~$D$%
-dimensional Gaussian wave packet with the same width $\sigma $ in all $D$
coordinate directions. However, $G_{\Sigma }(x)$ of Eq.~(\ref{eq:Weyl
Gaussian}) is, in general, not required to be a Wigner transform of
any~physical quantum state. Most importantly, $G_{\Sigma }(x)$ can be
arbitrarily narrow both in position and momentum, and hence does not have to
satisfy the Heisenberg uncertainty principle.\cite{Heller:1991}

Employing sufficiently narrow Gaussian functions (\ref{eq:Weyl Gaussian})
with fixed $\Sigma $ as our cells, the Wigner transform of a~general state
can be expanded as 
\begin{align}
\rho _{W}(x)& \equiv (C_{\Sigma }^{\rho _{W}}\!\ast G_{\Sigma })(x)  \notag
\\
& :=h^{-D}\int \!dz\,C_{\Sigma }^{\rho _{W}}\!(z)\,G_{\Sigma }(x-z),
\label{eq:Gaussian diag expansion}
\end{align}%
where the asterisk denotes the convolution of $G_{\Sigma }$ with $C_{\Sigma
}^{\rho _{W}}$. Function $C_{\Sigma }^{\rho _{W}}$, playing a role of
\textquotedblleft continuous expansion coefficient,\textquotedblright\ is
known as the \emph{inverse Weierstrass transformation} of $\rho _{W}$.\cite%
{Widder:1954} Thanks to normalization of $\rho _{W}$ and $G_{\Sigma }$,
integrating Eq.~(\ref{eq:Gaussian diag expansion}) over $x$ implies that $%
C_{\Sigma }^{\rho _{W}}$ is also normalized: $h^{-D}\int \!dz\,C_{\Sigma
}^{\rho _{W}}\!(z)=1$.

Equation~(\ref{eq:Gaussian diag expansion}) can be inverted via the
convolution theorem to obtain 
\begin{align}
C_{\Sigma }^{\rho_{W}}(z)& =\mathcal{F}^{-1}[\mathcal{F}[\rho_{W}]/\mathcal{F%
}[{G}_{\Sigma }]]  \notag \\
& \equiv h^{-D}\int \!d\eta \,e^{\eta ^{\mathsf{T}}\cdot \Sigma ^{-1}\cdot {%
\eta }/2\hbar ^{2}}e^{iz^{\mathsf{T}}\cdot {\eta }/\hbar }\mathcal{F}[\rho
_{W}]({\eta }),  \label{eq:weierstrass trans}
\end{align}%
where $\mathcal{F}[\cdot ]$ denotes the phase-space Fourier transform,%
\begin{equation}
\mathcal{F}[{\rho }_{W}]({\eta }):=h^{-D}\int \!dx\,\rho_{W}(x)\,e^{-ix^{%
\mathsf{T}}\cdot \eta /\hbar },
\end{equation}%
while $\mathcal{F}^{-1}[\cdot ]$ stands for its inverse. The Fourier
transform of $G_{\Sigma }$ can be evaluated analytically as 
\begin{equation}
\mathcal{F}[G_{\Sigma }](\eta )=e^{-\eta ^{\mathsf{T}}\cdot \Sigma
^{-1}\cdot \eta /2\hbar ^{2}}.
\end{equation}%
From Eq.~(\ref{eq:weierstrass trans}) we see that $C_{\Sigma }^{\rho
_{W}}(z) $ is well-defined only if $\mathcal{F}[{\rho }_{W}]$ decays
sufficiently faster than $\mathcal{F}[G_{\Sigma }]$. In other words, the
Gaussian cells must be sufficiently narrow in order that the integral (\ref%
{eq:weierstrass trans}) over $\eta $ converges.

If the initial state is a~Gaussian, i.e., $\rho_{W}(x)=G_{\Sigma
^{0}}(x-z^{0})$, the cell functions $G_{\Sigma }$ in Eq.~(\ref{eq:Gaussian
diag expansion}) can be conveniently chosen as scaled versions of $G_{\Sigma
^{0}}$ with widths in all coordinate and momentum directions multiplied by a
factor $\lambda $, where $0<\lambda \leq 1$, which is equivalent to setting $%
\Sigma =\Sigma ^{0}/\lambda ^{2}$. The width of cell $G_{\Sigma }$ may vary
from zero (a delta function) for $\lambda =0$ to the width of the initial
state $G_{\Sigma _{0}}$ for $\lambda =1$. The inverse Weierstrass transform (%
\ref{eq:weierstrass trans}) can be evaluated analytically for all admissible 
$\lambda $ (i.e., $0\leq \lambda \leq 1$) as 
\begin{equation}
C_{\Sigma }^{\rho_{W}}(z)=G_{\Lambda }(z-z^{0}),  \label{eq:weier_scaled}
\end{equation}%
where 
\begin{equation}
\Lambda =(1-\lambda ^{2})^{-1}\Sigma ^{0}=\lambda ^{2}(1-\lambda
^{2})^{-1}\Sigma .  \label{eq:Lambda}
\end{equation}%
Note that for $\lambda >1$ the inverse Weierstrass transform (\ref%
{eq:weierstrass trans}) diverges. The limiting cases of the sampling weight~(%
\ref{eq:weier_scaled}) are 
\begin{equation}
C_{\Sigma }^{\rho_{W}}(z)=G_{\Lambda }(z-z^{0})\rightarrow\left\{ 
\begin{array}{ll}
h^{D}\delta (z-z^{0}), & \lambda =1, \\[1.5ex] 
G_{\Sigma ^{0}}(z-z^{0}), & \lambda \rightarrow 0^{+},%
\end{array}%
\right.  \label{eq:Weierstrass for Gaussian}
\end{equation}%
and are represented, respectively, by the red dot at the top and red circle
at the bottom of Fig.~\ref{fig:cellularization}(b). Indeed, for $\lambda =1$%
, there is no freedom in the choice of the center of the single cell,
whereas in the limit $\lambda \rightarrow 0$, the sampling weight converges
to $\rho_{W}$.

Inserting the cellular expansion (\ref{eq:Gaussian diag expansion}) into the
DR formula~(\ref{eq:DR formula}) yields 
\begin{equation}
f_{\text{DR}}(t)=h^{-2D}\hspace{-0.1cm}\int \hspace{-0.1cm}dz^{0}\,C_{\Sigma
}^{\rho_{W}}(z^{0})\int \hspace{-0.1cm}dx^{0}\,G_{\Sigma
}(x^{0}-z^{0})e^{iS_{\text{DR}}(x^{0},t)/\hbar }.  \label{eq:DR with weiss}
\end{equation}%
In order to carry out the integration over $x^{0}$ analytically, one expands
the DR phase about point $z^{0}$ as $S_{\text{DR}}(x^{0},t)\approx S_{\text{%
CDR}}(x^{0},t;z^{0})$, where the CDR action is 
\begin{equation}
S_{\text{CDR}}(x^{0},t;z^{0}):=S_{\text{DR}}(z^{0},t)+\delta x^{\mathsf{T}%
}\cdot \alpha _{z^{0}}^{t}+\delta x^{\mathsf{T}}\cdot B_{z^{0}}^{t}\cdot
\delta x.  \label{eq:S_CDR}
\end{equation}%
In the last equation, $\delta x:=x^{0}-{\ z^{0}}$, $\alpha
_{z^{0}}^{t}:=\partial S_{\text{DR}}(z^{0})/\partial z^{0}$ is the gradient
of $S_{\text{DR}}$ at $z^{0}$, and $B_{z^{0}}^{t}$, already defined in Eq.~(%
\ref{eq:dS_expansion_B}), is, up to a factor $1/2$, the Hessian of $S_{\text{%
DR}}$ at $z^{0}$. Using the quadratic expansion (\ref{eq:S_CDR}), the
integral over $x^{0}$ in the double integral representation (\ref{eq:DR with
weiss}) of the DR is performed analytically to yield the final result---CDR: 
\begin{equation}
f_{\text{CDR}}(t)=\left\langle A_{\text{CDR}}(z^{0},t)e^{iS_{\text{DR}%
}(z^{0},t)/\hbar }\right\rangle _{C_{\Sigma }^{\rho_{W}}\!(z^{0})},
\label{eq:CDR formula}
\end{equation}%
with 
\begin{align}
A_{\text{CDR}}(z^{0},t)& :=\left\vert \det (\Sigma \cdot
K_{z^{0}}^{t})\right\vert ^{1/2}e^{-(\alpha _{z^{0}}^{t})^{\mathsf{T}}\cdot
K_{z^{0}}^{t}\cdot \alpha _{z^{0}}^{t}/2\hbar ^{2}},  \label{eq:A_CDR} \\
\Sigma \cdot K_{z^{0}}^{t}& =(I-{2i}B_{z^{0}}^{t}\cdot \Sigma ^{-1}/\hbar
)^{-1}.  \label{eq:Sigma_K}
\end{align}

Straightforward numerical implementation evaluates $f_{\text{CDR}}(t)$ in
Eq.~(\ref{eq:CDR formula}) by Monte Carlo importance sampling. This means
arithmetically averaging the estimator $A_{\text{CDR}}\exp (iS_{\text{DR}%
}/\hbar )$ over the set of $N$ initial conditions sampled from the weight $%
C_{\Sigma }^{\rho_{W}}\!$ using the Box-Muller algorithm for Gaussian
initial states or Metropolis algorithm for general states. [The positivity
of $C_{\Sigma }^{\rho_{W}}$ is for Gaussian initial states guaranteed by
Eq.~(\ref{eq:weier_scaled}).] Equivalently, one can think of this procedure
as expanding the Wigner transform $\rho_{W}$ of the initial state into a
finite set of Gaussians, i.e., 
\begin{equation}
\rho_{W}(x)\approx \sum_{n=1}^{N}C_{n}\,G_{\Sigma }(x-z_{n}),
\label{eq:rho_W=sum}
\end{equation}%
where $C_{n}=1/N$ and centers $\{z_{n}\}$ are sampled from $C_{\Sigma
}^{\rho_{W}}\!(z)$. This expansion is then combined with the quadratic
expansion (\ref{eq:S_CDR}) of $S_{\text{DR}}$ and substituted into the DR
formula (\ref{eq:DR formula integral}).

As mentioned above,~a natural value of the scaling parameter is $\lambda
=N^{-1/2D}$ for which the $N$ cells $G_{\Sigma ^{0}/\lambda ^{2}}$ cover
essentially the same phase-space volume as the initial state $\rho
_{W}(x)=G_{\Sigma ^{0}}(x-z^{0})$. Moreover, for $N=1$, Eq.~(\ref%
{eq:lambda_scaling}) gives $\lambda =1$. From Eq.~(\ref{eq:Weierstrass for
Gaussian}) we see that $C_{\Sigma }^{\rho _{W}}\!(z)$ degenerates to a~delta
function and the~single cell is identical to $\rho _{W}$. On the other hand, 
$N\rightarrow \infty $ implies $\lambda \rightarrow 0^{+}$ and Eq.~(\ref%
{eq:Sigma_K}) yields $\Sigma \cdot K_{z^{0}}^{t}\rightarrow I$ and $%
K_{z^{0}}^{t}\rightarrow 0.$ Since for $\lambda \rightarrow 0^{+}$, $%
C_{\Sigma }^{\rho _{W}}(z)\rightarrow G_{\Sigma ^{0}}(z-z^{0})=\rho _{W}(z)$
and $A_{\text{CDR}}\rightarrow 1$, comparison of Eqs.~(\ref{eq:CDR formula})
and (\ref{eq:DR formula}) confirms that the CDR reduces in the limit $%
N\rightarrow \infty $ to the original DR, as promised. Note that this
desirable property was satisfied neither by the original CDR nor by standard
cellularization or Filinov filtering procedures for the Van Vleck or
Herman-Kluk propagators.

Several further improvements are possible: First, a~significant boost in
computational efficiency could be gained with ideas implemented in the \emph{%
generalized} Filinov filtering\cite{Makri_Miller:1987,Wang:2001} or
stationary phase Monte Carlo method.\cite{Doll:1988} Motivated by the
generalized Filinov method, for instance, one would add a complex linear
term to the exponent of the Gaussian cell to ensure that the overall phase
of the integrand of the $x^{0}$ integral in Eq.~(\ref{eq:DR with weiss})
were approximately stationary, making the original integral more amenable to
Monte Carlo integration. This is in contrast to the original Filinov
approach,\cite{Filinov:1986} which does not employ an additional phase.
Another improvement relies on Sobol sampling,\cite%
{Press_NumericalRecipes:2010} which actively seeks different initial
conditions while preserving the normal distribution, and was used, e.g., by
Walton and Manolopoulos.\cite{Walton_Manolopoulos:1996} Finally, it is
advantageous to allow the expansion coefficients $C_{n}$ in Eq.~(\ref%
{eq:rho_W=sum}) to differ from $1/N$. Specifically, one finds the optimal
coefficients $C_{n}$ for given, already sampled, Gaussian centers $\{z_{n}\}$
by minimizing the residual $L^{2}$ error of the expansion~(\ref{eq:rho_W=sum}%
) under the constraints 
\begin{subequations}
\label{eq:def_Cn}
\begin{align}
\quad & \sum_{n=1}^{N}C_{n}=1\text{ and}  \label{eq:def_Cn_i} \\
\quad & C_{n}\geq 0,\text{ }n=1,\ldots ,N,  \label{eq:def_Cn_ii}
\end{align}%
which guarantee that $f_{\text{CDR}}(0)=1$ and $|f_{\text{CDR}}(t)|\leq 1$.
From numerical point of view, this amounts to solving a~convex quadratic
program.\cite{book_Nocedal} As demonstrated in Sec.~\ref{sec:Examples}, this
procedure further enhances efficiency, nevertheless the acceleration due to
the cellularization procedure itself is dominant. 

In practice, one should always use all five \textquotedblleft
tricks,\textquotedblright\ i.e., sampling (\ref{eq:CDR formula}) from the
inverse Weierstrass transform, scaling (\ref{eq:lambda_scaling}) of the
cells with $N$, generalized Filinov filtering,\cite%
{Makri_Miller:1987,Wang:2001} Sobol sampling,\cite%
{Press_NumericalRecipes:2010} and optimal coefficients (\ref{eq:def_Cn_i})-(%
\ref{eq:def_Cn_ii}). Although clearly beneficial, generalized Filinov
filtering and Sobol sampling were not employed here, in order to clearly
separate the effect of the three new ideas presented: sampling (\ref{eq:CDR
formula}) from the inverse Weierstrass transform, scaling (\ref%
{eq:lambda_scaling}) of the cells with $N$, and optimal coefficients (\ref%
{eq:def_Cn_i})-(\ref{eq:def_Cn_ii}).

\subsection{Cellular DR with prefactor correction}

The numerical prerequisites of the CDR (Subsec.~\ref{subsec:CDR}) and DRP
(Subsec.~\ref{subsec:DRP}) are the same---the cost per trajectory is
determined by evaluating the Hessian of $S_{\text{DR}}$ with respect to
initial conditions. This allows for a~straightforward combination of the
methods, without increasing the cost per trajectory, by multiplying the
contribution~(\ref{eq:CDR formula}) of each trajectory with the prefactor (%
\ref{eq:A_DRP}) and thus obtaining the \textit{cellular dephasing
representation with prefactor} (CDRP): 
\end{subequations}
\begin{equation}
f_{\text{CDRP}}(t)=\left\langle A_{\text{CDRP}}(z^{0},t)e^{iS_{\text{DR}%
}(z^{0},t)/\hbar }\right\rangle _{C_{\Sigma }^{\rho_{W}}(z^{0})},
\label{eq:CDRP formula}
\end{equation}%
where 
\begin{equation}
A_{\text{CDRP}}(z^{0},t):=A_{\text{DRP}}(z^{0},t)A_{\text{CDR}}(z^{0},t).
\label{eq:A_CDRP}
\end{equation}%
In principle, the CDRP should benefit both from the enhanced efficiency of
the CDR and improved accuracy of the DRP, as depicted in Fig.~\ref{fig:CDRP
diagram}.

\begin{figure}[htbp]
\includegraphics[width=0.9\columnwidth]{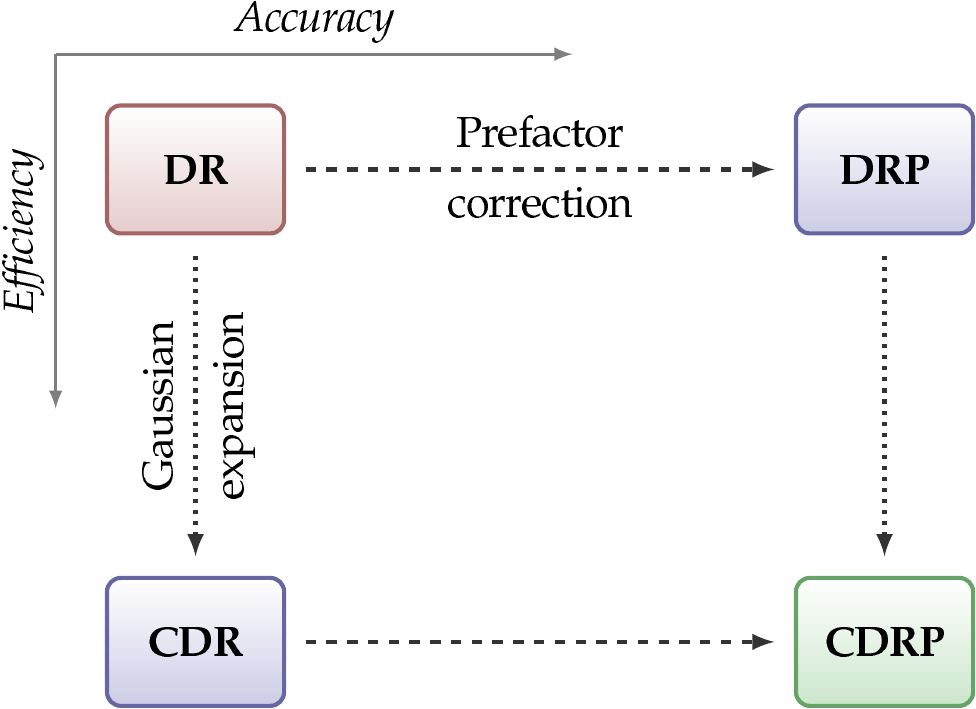} 
\caption{Relations between several approximations for time correlation function
(\ref{eq:fid_amplitude}). Typically, the accuracy increases along the horizontal arrows, corresponding 
to adding the prefactor (\ref{eq:DRP formula}), while the efficiency improves in the downward direction, 
corresponding to the cellularization procedure (\ref{eq:CDR formula}).
} \label{fig:CDRP diagram}
\end{figure}

As for the asymptotic computational complexity of Eq.~(\ref{eq:CDRP formula}%
) per trajectory, a~straightforward implementation scales with system's
dimensionality $D$ and total propagation time $t$ as $\mathcal{O}(D^{3}t)$.
Linear scaling with time is easily verified by direct inspection of Eq.~(\ref%
{eq:CDRP formula}), while the cubic dependence on $D$ is due to the
necessity to propagate the stability matrix and due to the matrix operations
implicit in Eqs.~(\ref{eq:CDR formula}) and (\ref{eq:A_CDR}). The CDRP is
thus cheaper than, e.g., the popular \textit{Forward Backward Initial Value
Representation}\cite{Sun_Miller:1999} which would scale as $\mathcal{O}%
(D^{3}t^{2})$. 

\section{Numerical Examples\label{sec:Examples}}

In this section we will show how the CDRP approximation improves the
accuracy of the time correlation function (\ref{eq:fid_amplitude}) and
stimulated emission spectrum (\ref{eq:sigma_wp}) for several well-known
systems.

\subsection{Harmonic oscillators\label{subsec:lho}}

As the first example we consider two quadratic Hamiltonians in $D$
dimensions: 
\begin{subequations}
\label{eq:H Harmonic}
\begin{align}
H_{g}& =\frac{1}{2}x^{\mathsf{T}}\cdot \mathcal{H}_{g}\cdot x, \\
H_{e}& =\frac{1}{2}(x-d)^{\mathsf{T}}\cdot \mathcal{H}_{e}\cdot (x-d)+V_{0},
\end{align}%
where $V_{0}$ is the gap between the two potential wells, 
\end{subequations}
\begin{equation}
\mathcal{H}_{j}:=\frac{\partial ^{2}H_{j}}{\partial x^{2}}=%
\begin{pmatrix}
k_{j} & 0_{D} \\ 
0_{D} & m^{-1}%
\end{pmatrix}%
\quad \quad (j=g,\,e),
\end{equation}%
is the $2D\times 2D$ Hessian matrix of $H_{j}$, $k_{j}$ being the
force-constant matrix, $m_{ij}=m_{i}\delta _{ij}$ is the $D\times D$ matrix
of masses, and $d=(d_{q},d_{p})$ is the phase-space displacement of the two
Hamiltonians: e.g., $d_{q}$ is the coordinate distance between the two
potential minima. The Hessian of the average Hamiltonian is given by the
(invertible) $2D\times 2D$ matrix 
\begin{equation}
\mathcal{H}:=\frac{\partial ^{2}H}{\partial x^{2}}=%
\begin{pmatrix}
k & {0}_{D} \\ 
0_{D} & m^{-1}%
\end{pmatrix}%
,
\end{equation}%
where $k:=(k_{g}+k_{e})/2$. The path driven by the average Hamiltonian is 
\begin{equation}
x^{t}={M}^{t}\cdot (x^{0}-\delta )+\delta ,
\end{equation}%
where $M^{t}:=\exp (t\,J\cdot \mathcal{H})$ is the stability matrix for $%
\mathcal{H}$ and $\delta :=\mathcal{H}^{-1}\cdot \mathcal{H}_{e}\cdot d/2$.
Since the Hamiltonians (\ref{eq:H Harmonic}) are quadratic, it is possible
to evaluate the DR phase analytically for an arbitrary initial condition $%
x^{0}$ as 
\begin{equation}
S_{\text{DR}}(x^{0},t)\equiv (x^{0}-\delta )^{\mathsf{T}}\cdot B^{t}\cdot
(x^{0}-\delta )+(x^{0}-\delta )^{\mathsf{T}}\cdot v^{t}+a^{t},
\label{eq:S_DR quadratic}
\end{equation}%
where 
\begin{align}
B^{t}& \equiv -\frac{1}{2}\int_{0}^{t}dt^{\prime}(M^{t^{\prime}})^{\mathsf{T}%
}\cdot \Delta \mathcal{H}\cdot M^{t^{\prime }}, \\
v^{t}& :=-2\int_{0}^{t}dt^{\prime}(M^{t^{\prime}})^{\mathsf{T}}\cdot \left( 
\mathcal{H}+\frac{\Delta \mathcal{H}}{2}\right) \cdot \delta , \\
a^{t}& :=\left( V_{0}+\frac{1}{2}\delta ^{\mathsf{T}}\cdot \Delta {\mathcal{H%
}}\cdot \delta _{+}\right) t  \label{eq:B quadratic}
\end{align}%
with $\Delta \mathcal{H}:=\mathcal{H}_{g}-\mathcal{H}_{e}$ and $\delta _{+}:=%
\mathcal{H}^{-1}\cdot \mathcal{H}_{g}\cdot d/2$. Note that in the harmonic
systems, the cellular schemes are exactly equal to their noncellular
analogs, e.g.,%
\begin{equation}
f_{\text{CDR}}(t)\equiv f_{\text{DR}}(t)=\left\langle e^{iS_{\text{DR}%
}(x^{0},t)/\hbar }\right\rangle _{\rho_{W}(x^{0})}.  \label{eq:DR SHO}
\end{equation}%
[However, if a discrete Gaussian expansion (\ref{eq:rho_W=sum}) is used, the
accuracy of the results will be limited by the error inherent in Eq.~(\ref%
{eq:rho_W=sum}).] Since $B^{t}$ and hence $A_{\text{DRP}}(t)$ are in this
case independent of $x^{0}$, the DRP and CDRP can be calculated for an
arbitrary initial state as 
\begin{equation}
f_{\text{CDRP}}(t)\equiv f_{\text{DRP}}(t)\equiv A_{\text{DRP}}(t)f_{\text{DR%
}}(t).  \label{eq:CDRP SHO}
\end{equation}

Explicit formulas for one degree of freedom are 
\begin{subequations}
\begin{align}
B^{t} & = -\Delta k 
\begin{pmatrix}
t+\sin(2\omega t)/2\omega & \sin^{2}(\omega t)/m\omega^{2} \\ 
\sin^{2}(\omega t)/m\omega^{2} & \frac{t}{(m\omega)^{2}}-\frac{\sin(2\omega
t)}{2\omega(m\omega)^{2}}%
\end{pmatrix}
, \\
v^{t} & =\left( 1-\frac{\Delta k}{2m\omega^{2}}\right) ^{2} 
\begin{pmatrix}
m\omega\sin(\omega t) \\ 
1-\cos(\omega t)%
\end{pmatrix}
, \\
a^{t} & = V_{o}t+\frac{d^{2}}{8}\Delta{k}\left[ 1-\left( \frac{\Delta k}{%
2m\omega^{2}}\right) ^{2}\right] t.
\end{align}
Here, $\omega^{2}:= k/m$, $\Delta k:=k_{g}-k_{e}$, and $d_{p}=0$, i.e., $d$
has only position components. Additionally, the determinant prefactor is
given by 
\end{subequations}
\begin{equation}
A_{\text{DRP}}(t)=\left| 1+\left( \frac{\Delta k}{4m\omega} \right) ^{2}
\left( t^{2}-\frac{\sin^{2}\omega t}{\omega^{2}}\right) \right| ^{\frac {1}{2%
}}.
\end{equation}

\begin{figure*}[htbp]
\includegraphics{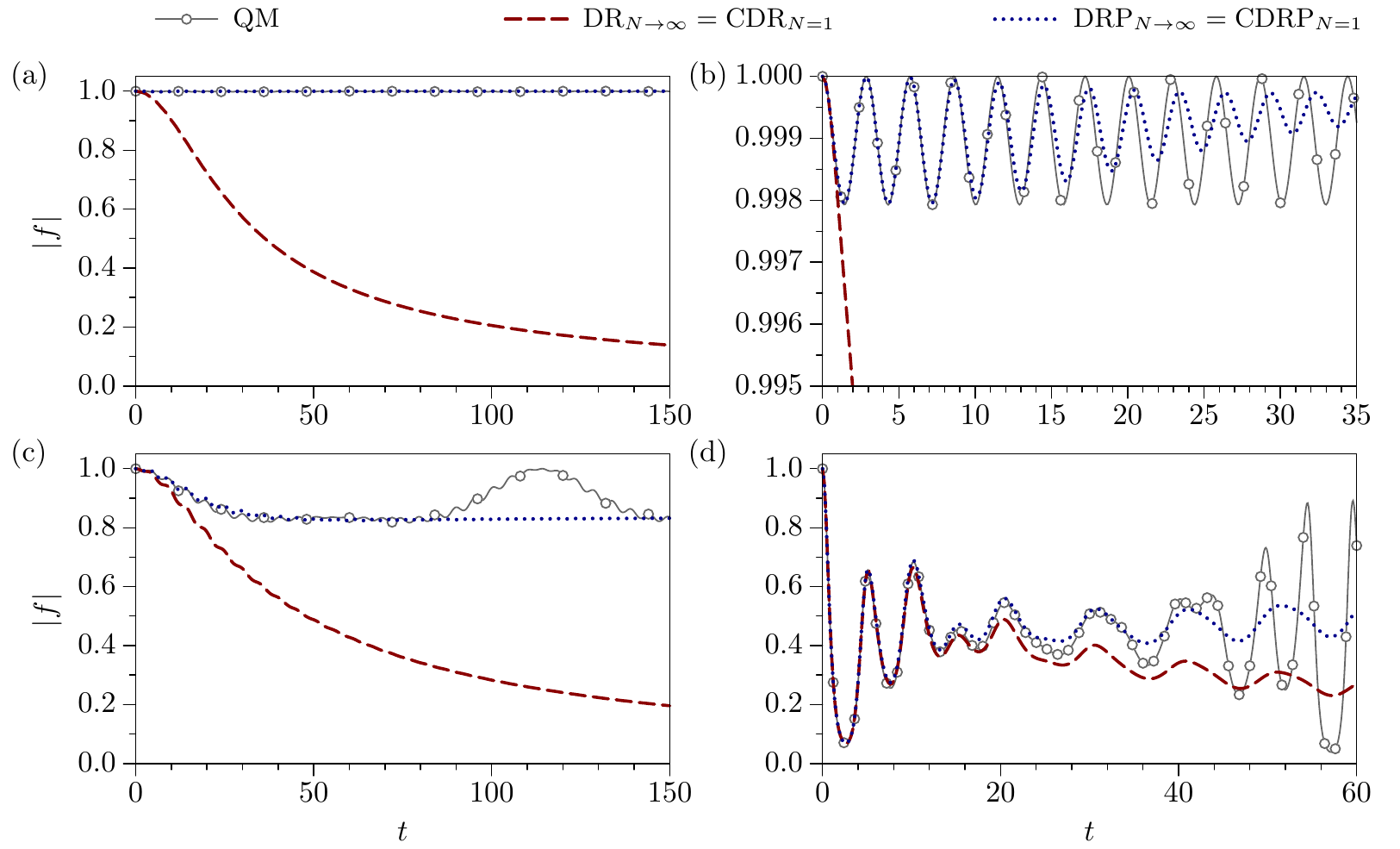}
\caption{Time correlation functions for time-resolved stimulated emission
spectrum with a zero time delay ($\protect\tau =0$) in a one-dimensional
harmonic potential (\protect\ref{eq:H Harmonic}) with $V_{0}=0$, force
constants $k_{g}=1$ and $k_{e}=1.2$. The initial state is a Gaussian wave
packet with width $\protect\sigma $ and centered at $z^{0}$. (a) $m=1$, $%
\protect\sigma =1$, $z^{0}=(0,0)$, and $d=(0,0)$. (b) Detail of panel (a).
(c) $m=3$, $\protect\sigma =1$, $z^{0}=(0.3,0.3)$, and $d=(0,0)$. (d) $m=3$, 
$\protect\sigma =2$, $z^{0}=(0.5,0)$, and $d=(1,0)$. }
\label{fig:SHO}
\end{figure*}

Figure~\ref{fig:SHO} shows the fully converged time correlation functions
for zero time delay in one-dimensional harmonic oscillator (\ref{eq:H
Harmonic}) using a Gaussian initial state. We observe the effect of the
prefactor (\ref{eq:A_DRP}): it enhances the accuracy compared with the DR,
so that the approximate time correlation function does not decay with
increasing time. Note that the Fourier transforms of time correlations shown
in Fig.~\ref{fig:SHO} can be interpreted both as TRSE spectra with zero time
delay and as \emph{continuous-wave absorption spectra}. 

\begin{figure}[htbp]
\includegraphics{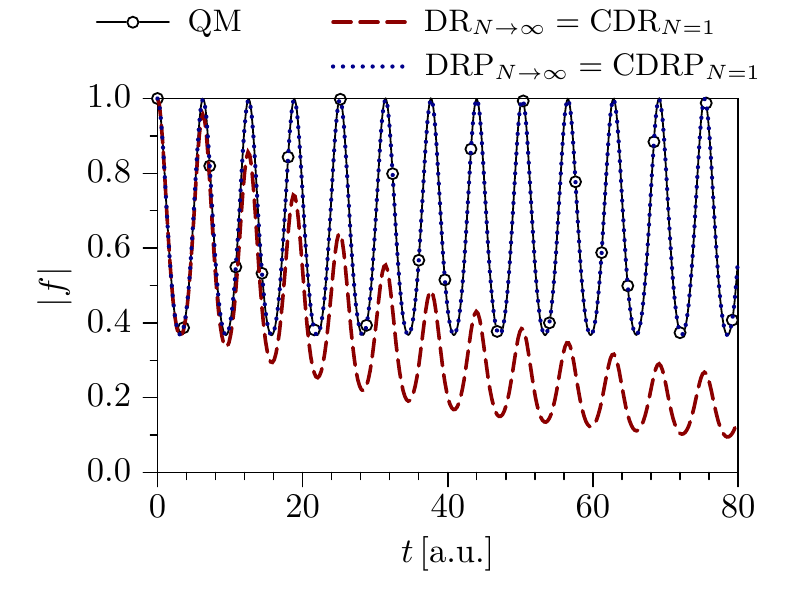} 
\caption{\label{fig:SHO2D}
Time correlation function for time-resolved stimulated emission spectrum in a two-dimensional harmonic oscillator model.
Displacements are $d=(d_q,d_p)$ with $d_q=(1,0)$ and $d_p=(0,0)$, $V_0=10$, and $m=1$, and force constants [according to  Eq.~(\ref{eq:SHO2D})]
are $k_1=1$ and $k_2=2$. The initial state is the ground state of the ground PES. Time delay $\tau=10$.
}
\end{figure}

Now we consider a two-dimensional harmonic system (\ref{eq:H Harmonic}) with 
$d_{q}=(d_{1},0)$, $d_{p}=(0,0)$, 
\begin{equation}
k_{g}=%
\begin{pmatrix}
k_{1} & 0 \\ 
0 & k_{1}%
\end{pmatrix}%
,\quad \text{and}\quad k_{e}=%
\begin{pmatrix}
k_{1} & 0 \\ 
0 & k_{2}%
\end{pmatrix}%
,  \label{eq:SHO2D}
\end{equation}%
which is a prototype of the breakdown of the DR in simple molecular systems.
While the DR describes exactly the behavior of the \textquotedblleft
excited\textquotedblright\ mode corresponding to displaced simple harmonic
oscillators,\cite{Mukamel:1982,*book_Mukamel} this agreement is lost due to
the decay of the DR in the \textquotedblleft silent\textquotedblright\ mode,
corresponding to harmonic oscillators with different force constants [as in
Fig.~\ref{fig:SHO}(a)], in which the DR breaks down. In other words, the
breakdown of the DR for the uninteresting mode covers up the accurate
information about the interesting mode. Figure~\ref{fig:SHO2D} shows the
time correlation function for time delay $\tau =10\text{, confirming that }$%
the DRP can in this system almost completely remove the error introduced by
the DR.

\subsection{\label{subsec:pyrazine}Pyrazine model}

The next system is based on the four-dimensional vibronic coupling model
taking into account normal modes $\nu _{1}$, $\nu _{\text{6a}}$, $\nu _{%
\text{9a}}$, and $\nu _{\text{10a}}$ of pyrazine.\cite{Stock_Woywod:1995} We
employ the $S_{0}$ and $S_{1}$ surfaces from Ref.~%
\onlinecite{Stock_Woywod:1995}, but disregard the nonadiabatic coupling
between states $S_{1}$ and $S_{2}$ since for the $S_{0}\rightarrow {}S_{1}$
excitation this coupling is much less important than for the $%
S_{0}\rightarrow {}S_{2}$ excitation and since nonadiabatic dynamics is not
our primary focus. However, even this simplified model requires a~nontrivial
Duschinsky rotation\cite{Duschinsky:1937,*Ozkan:1990} connecting normal
modes of the $S_{0}$ and $S_{1}$ states.

\begin{figure}[htbp]
\includegraphics{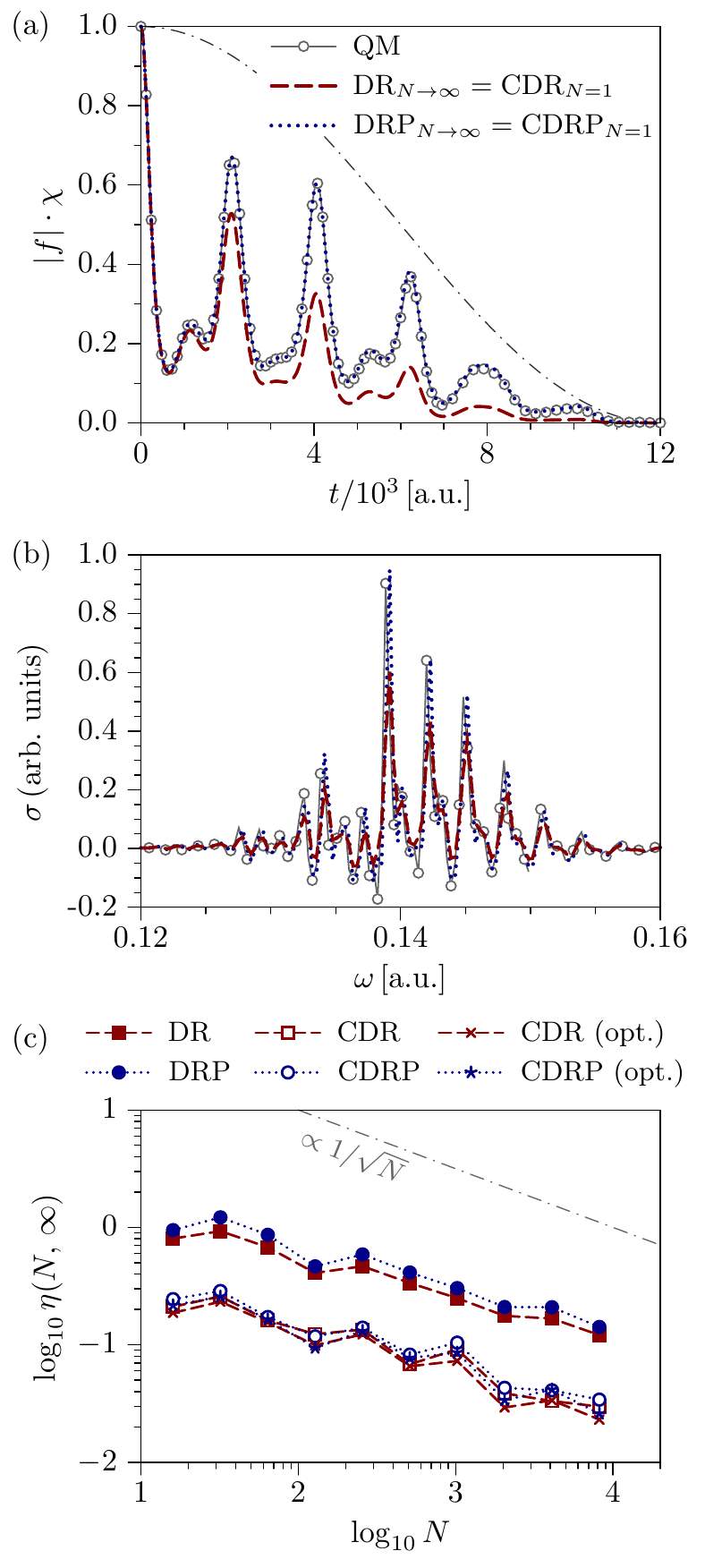} 
\caption{Time-resolved stimulated emission in the pyrazine model of Subsec.~\ref{subsec:pyrazine}.
Initial state is the ground state of the $S_0$ surface, the delay
time $\tau=2\times10^3\,\text{a.u.}\approx48\,\text{fs}$.
(a) Time correlation function [damped by $\chi(t)$ of Eq.~(\ref{eq:damping}), shown as a dash-dotted line].
(b) Corresponding spectrum.
(c) Convergence error $\eta$ [defined in Eq.~(\ref{eq:err_eta})] of the damped correlation function as a~function of the number
of trajectories $N$. The points labeled by ``opt.'' were computed with optimized
expansion coefficients $C_n$ of Eq.~(\ref{eq:def_Cn}) (see Subsec.~\ref{subsec:CDR}).
\label{fig:pyrazine}}
\end{figure}

Since the pyrazine model is globally quadratic, the action expansion in Eq.~(%
\ref{eq:S_CDR}) is exact (as discussed in Subsec.~\ref{subsec:lho}) and thus
the fully converged DR and DRP correlation functions can be obtained by the
cellular variants CDR$_{N=1}$ and CDRP$_{N=1}$ of these methods obtained
with a~single trajectory.

Figure~\ref{fig:pyrazine}(a) shows pyrazine TRSE correlation function $%
f(t,\tau )$, calculated for a particular delay time $\tau \approx 48$~fs and
multiplied by a~phenomenological damping function\cite{book_MCTDH} 
\begin{equation}
\chi (t):=\cos ^{2}[{\pi t}/(2T)]\,\theta (T-t),  \label{eq:damping}
\end{equation}%
where $T$ denotes the total propagation time. Parameters of the calculation
are summarized in the caption of Fig.~\ref{fig:pyrazine}. The DRP is shown
in Fig.~\ref{fig:pyrazine}(a) to yield an excellent agreement with the
quantum calculation. This is also confirmed in the corresponding spectrum
[Fig.~\ref{fig:pyrazine}(b)], computed as the Fourier transform (\ref%
{eq:sigma_wp}) of the damped correlation function.

Finally, Fig.~\ref{fig:pyrazine}(c) compares the convergence behavior of
individual methods. The convergence is quantified by the relative $L^{2}$
error achieved for $N\ll N_{\text{ref}}$ trajectories: 
\begin{equation}
\eta (N,N_{\text{ref}}):=\lVert (f_{N}-f_{N_{\text{ref}}})\,\chi \rVert
/\lVert f_{N_{\text{ref}}}\,\chi \rVert ,  \label{eq:err_eta}
\end{equation}%
where $\lVert f\rVert ^{2}:=\int_{0}^{T}\!d\tau ^{\prime }\lvert f(\tau
^{\prime })\rvert ^{2}$. The subscript $N$ of $f_{N}$ in Eq.~(\ref%
{eq:err_eta}) emphasizes that the quantity $f_{N}$ was computed with $N$
trajectories, while the fully converged results are denoted by $N\rightarrow
\infty $. Time integrals appearing implicitly in Eq.~(\ref{eq:err_eta}) are
evaluated with Simpson's method. 
The cellularization accelerates convergence by lowering the number of
trajectories required to achieve the same statistical error by about two
orders of magnitude [Fig.~\ref{fig:pyrazine}(c)]. Additional minor
improvement is achieved by optimizing the expansion coefficients in Eq.~(\ref%
{eq:rho_W=sum}) using constraints~(\ref{eq:def_Cn}). 

\subsection{\label{subsec:quartic}Quartic oscillator}

After discussing harmonic systems, which are rather simple even in high
dimensions, let us turn to the opposite limit of chaotic dynamics, which can
present difficulties even in few dimensions. In particular, we consider a
two-dimensional chaotic quartic oscillator.\cite%
{Meyer:1986,*Waterland_Yuan:1988,*Eckhardt_Hose:1989,*Bohigas_Tomsovic:1993,*Revuelta_Vergini:2012}
The two potential energy surfaces, 
\begin{equation}
V_{j}(q_{1},q_{2})=\frac{q_{1}^{2}q_{2}^{2}}{2}+\frac{\beta _{j}}{4}%
(q_{1}^{4}+q_{2}^{4}),  \label{eq:pot_quartic}
\end{equation}%
differ only in the parameter $\beta _{j}>0$. Chaotic behavior is due to the
coupling term $q_{1}^{2}q_{2}^{2}/2$ since in the limit $\beta
_{j}\rightarrow \infty $, the Hamiltonian $T+V_{j}$ becomes separable and
hence integrable.

\begin{figure}[htbp]
\includegraphics{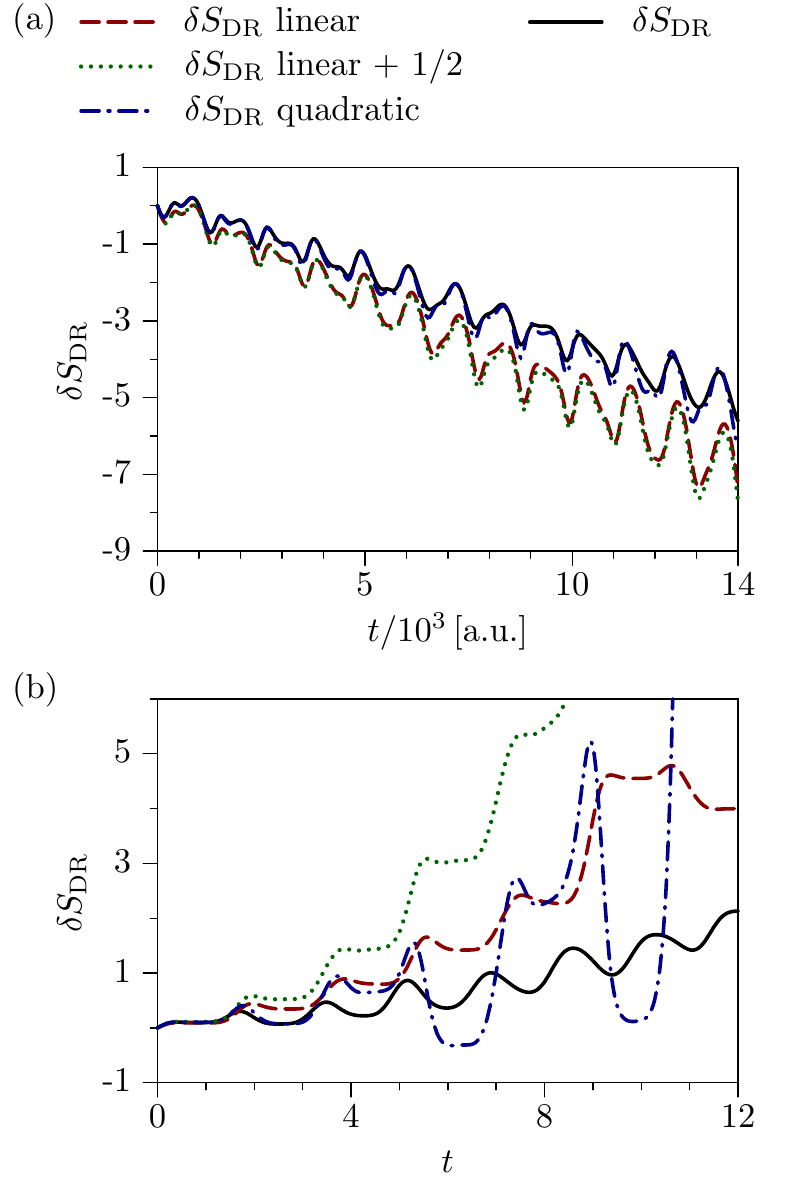} 
\caption{\label{fig:ds}
Time dependence of the DR action difference $\delta S_{\text{DR}}$ of Eq.~(\ref{eq:delta_dS})
calculated for two neighboring trajectories (initial conditions $z^0$ and $w^0$).
(a) Quartic oscillator (Fig.~\ref{fig:quartic}). (b)  Collinear NCO molecule (Fig.~\ref{fig:nco}).
Delay times are as in Figs.~\ref{fig:quartic} and \ref{fig:nco}, $z^0$ is the phase-space center of the
Gaussian initial state (of width $\sigma$) and $w^0=z^0+(\sigma,0)/2$.
The order of the expansion (\ref{eq:S_CDR}) is distinguished by line type: ``linear'' (dashed),
``quadratic'' (dotted), and ``linear + $1/2$'' (dash-dotted). The symbol $1/2$ signifies that
the derivatives of the stability matrix in Eq.~(\ref{eq:dS_hessian}) are neglected.
Solid line shows numerically exact $\delta S_{\text{DR}}$.
}
\end{figure}

Due to the chaotic character of this system, one expects that the central
ingredient of the cellularization, i.e., the quadratic expansion of the
action difference in Eq.~(\ref{eq:S_CDR}) will be poor and hinder
convergence. This is indeed confirmed in Fig.~\ref{fig:ds}(a), showing the
difference of the DR action (\ref{eq:DR phase}) for two neighboring
trajectories specified by initial conditions $z^{0}$ and $w^{0}$, i.e., 
\begin{equation}
\delta S_{\text{DR}}(t):=S_{\text{DR}}(w^{0},t)-S_{\text{DR}}(z^{0},t).
\label{eq:delta_dS}
\end{equation}%
This quantity is then compared with predictions based on the quadratic
expansion (\ref{eq:S_CDR}) and its linear part. The expansion order denoted
\textquotedblleft linear + $1/2$\textquotedblright\ is a widely used
approximation\cite{Heller:1991,Walton_Manolopoulos:1996} to the quadratic
expansion (\ref{eq:S_CDR}) within which one neglects the third derivatives
of the potential (see Appendix~\ref{ap:Derivatives}). Figure~\ref{fig:ds}(a)
shows clearly that in the quartic oscillator the quadratic expansion (\ref%
{eq:S_CDR}) is reliable only for short times and that the linear expansion
is superior to the presumably more accurate \textquotedblleft linear + $1/2$%
\textquotedblright\ approach.

\begin{figure*}[htbp]
\includegraphics{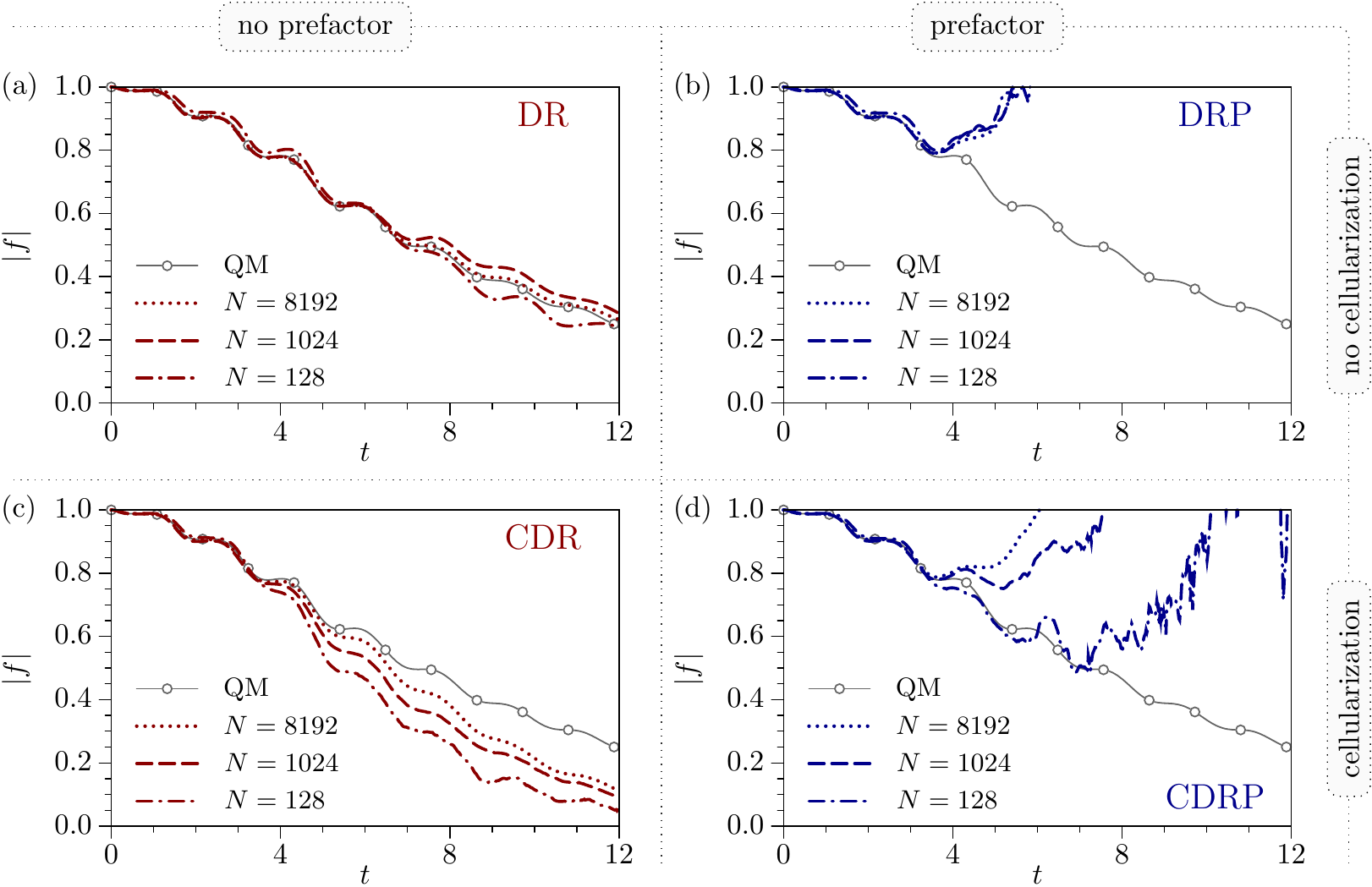}
\caption{Time correlation function for time-resolved stimulated emission in
quartic oscillator (\protect\ref{eq:pot_quartic}) corresponding to zero time
delay. Initial state is a~Gaussian wave packet [Eq.~(\protect\ref{eq:Weyl
Gaussian})] with $\protect\sigma _{1}=\protect\sigma _{2}=1$ centered at $%
(Q_{\text{init}},P_{\text{init}})$, where $Q_{\text{init}}=(0,4)$ and $P_{%
\text{init}}=(4,0)$. Masses $m_{1}=m_{2}=1$ and the potential energy
surfaces (\protect\ref{eq:pot_quartic}) are specified by $\protect\beta %
_{0}=0.2$ and $\protect\beta _{1}=0.2125$. }
\label{fig:quartic}
\end{figure*}

As a~consequence, Fig.~\ref{fig:quartic}, comparing the TRSE correlation
functions, shows that the method of choice for the quartic oscillator is the
\textquotedblleft bare\textquotedblright\ DR [Fig.~\ref{fig:quartic}(a)],
since the CDR [Fig.~\ref{fig:quartic}(c)] converges more slowly, while the
DRP and CDRP are reliable only for short times since the prefactor (\ref%
{eq:A_DRP}) (understood as a~function of time for fixed initial conditions)
grows quickly and oscillates widely at later times.

\subsection{\label{subsec:nco}Collinear NCO molecule}

Typical chemical systems are neither globally harmonic as our pyrazine-based
model, nor---fortunately---as strongly chaotic as the quartic oscillator. In
our last example we therefore consider a realistic, anharmonic system, in
order to see how the CDR, DRP, and CDRP might perform in typical situations.
For this purpose, we chose a~two-dimensional model of the collinear NCO
molecule based on the $\text{X}^{2}\Pi $ (ground) and $\text{A}^{2}\Sigma
^{+}$ (excited) PESs.\cite{Li_Carter:1993} The PESs are given in Ref.~%
\onlinecite{Li_Carter:1993} in a~form of a~polynomial fitted to \textit{ab
initio} calculations on the domain $r_{1,2}\in \lbrack 2,2.6]\,$a.u. and $%
\theta \in \lbrack 152^{\circ },208^{\circ }]$, specified in $r_{1}$ (N--C), 
$r_{2}$ (C--O) bond-length coordinates and the bond angle $\theta $. We set $%
\theta =\pi $ (equilibrium value) and describe the reduced two-dimensional
surfaces in the $r_{1}$ and $r_{2}$ coordinates by a~simplified two-term
form 
\begin{equation}
V(r_{1},r_{2})=V_{0}+\sum_{j=1,2}\!\!D_{j}\left\{ 1-\exp [-\beta
_{j}(r_{j}-r_{j}^{\text{e}})]\right\} ^{2}\!,  \label{eq:nco_pot}
\end{equation}%
where the equilibrium bond lengths $r_{j}^{\text{e}}$ are the same as in
Ref.~\onlinecite{Li_Carter:1993}, while the parameters $V_{0}$, $D_{1,2}$,
and $\beta _{1,2}$ were obtained by fitting potential (\ref{eq:nco_pot}) to
the functional form of Ref.~\onlinecite{Li_Carter:1993} on the domain $%
r_{j}\in \lbrack r_{j}^{\text{e}}-\delta ,r_{j}^{\text{e}}+\delta ]$ with $%
\delta =0.15\,$a.u. Resulting values are summarized in Tab.~\ref%
{tab:nco_pot_params}. These parameters differ from the values employed in
our earlier work\cite{Wehrle_Sulc:2011,Sulc_Vanicek:2012} and should better
reflect the dynamics of this system. Frequency-mass-scaled normal mode
coordinates of the $\text{X}^{2}\Pi $ PES were used so that the vibrational
ground state is in the harmonic approximation described by a~Gaussian with
unit widths centered at the origin. 

\begin{table}[htp]
\renewcommand{\arraystretch}{1.3} 
\caption{\label{tab:nco_pot_params}Parameters\footnote{All
quantities are given in atomic units.} of the collinear NCO model
(\ref{eq:nco_pot}).}. 
\begin{ruledtabular}
\begin{tabular}{ld{4.3}d{1.3}d{1.3}d{1.3}d{1.3}d{1.3}d{1.3}}
&\multicolumn{1}{c}{$V_0$}&\multicolumn{1}{c}{$D_1$}&\multicolumn{1}{c}{$\beta_1$}&\multicolumn{1}{c}{$r_1^{\text{e}}$}&\multicolumn{1}{c}{$D_2$}&\multicolumn{1}{c}{$\beta_2$}&\multicolumn{1}{c}{$r_2^{\text{e}}$}\\\hline
$\text{X}^{2}\Pi$&-167.653&0.150&1.698&2.302&0.333&1.160&2.246\\
$\text{A}^{2}\Sigma^{+}$&-167.549&0.144&1.984&2.234&0.398&1.140&2.232\\
\end{tabular}
\end{ruledtabular}
\end{table}


The initial state for the TRSE calculation was prepared by the following
procedure.\cite{Baer_Kosloff:1995} First, we computed the $\text{X}^{2}\Pi $
ground vibrational state by imaginary-time propagation. This state was then
pumped to the $\text{A}^{2}\Sigma ^{+}$ PES, propagated there for a net time
of $520\,\text{a.u.}\approx 12.6\,\text{fs}$, dumped to $\text{X}^{2}\Pi $,
and propagated for additional $480\,\text{a.u.}\approx 11.6\,\text{fs}$. In
order to facilitate computation of $C_{\Sigma }^{\rho_{W}}(z)$ in Eq.~(\ref%
{eq:weierstrass trans}), we approximated the resulting state by a~single
Gaussian. An independent quantum calculation confirmed that this does not
impact the spectrum significantly.

\begin{figure}[htbp]
\includegraphics{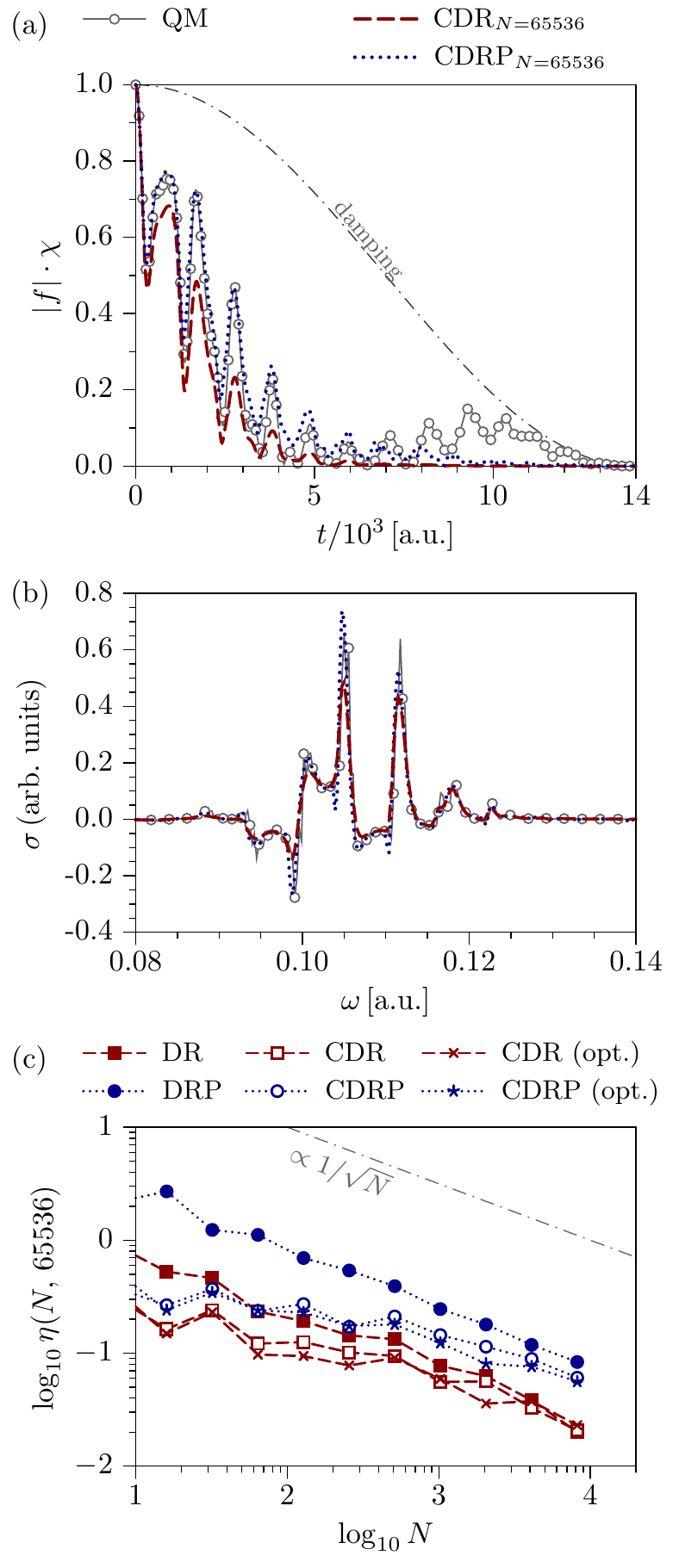} 
\caption{
Time-resolved stimulated emission in the NCO model of Subsec.~\protect\ref{subsec:nco}.
Initial state is a~non-stationary state prepared by a~pump-dump procedure\protect\cite{Baer_Kosloff:1995,Sulc_Vanicek:2012}
discussed in the text, the delay time $\tau=1200\,$a.u. $\approx$ $29\,$fs.
(a) Time correlation function [damped by $\chi(t)$ of Eq.~(\ref{eq:damping}), shown as a dash-dotted line].
(b) Corresponding spectrum.
(c) Convergence error $\eta$ [defined in Eq.~(\ref{eq:err_eta})] of the damped correlation
function as a~function of the number of trajectories $N$. The points labeled by
``opt.'' were computed with optimized expansion
coefficients $C_n$ of Eq.~(\ref{eq:def_Cn}) (see Subsec.~\ref{subsec:CDR}).
\label{fig:nco}}
\end{figure}

The TRSE correlation function for a delay time of $29$~fs is displayed in
Fig.~\ref{fig:nco}(a), confirming that the prefactor correction extends the
agreement of the DR with the quantum correlation function to longer times.
As a consequence, the prefactor correction yields sharper peaks in the
corresponding spectrum, shown in Fig.~\ref{fig:nco}(b). Finally, Fig.~\ref%
{fig:nco}(c), comparing the statistical convergence of the DR, CDR, DRP, and
CDRP, confirms that in NCO the cellularization increases numerical
efficiency, although the effect is---as expected---smaller than in the
harmonic pyrazine model [Fig.~\ref{fig:pyrazine}(c)].

\subsection{\label{subsec:comp_details}Computational details}

Classical trajectories needed in the DR, CDR, DRP, and CDRP were calculated
with a fourth-order symplectic integrator, while quantum calculations
employed the corresponding fourth-order split-operator method.\cite%
{Wehrle_Sulc:2011} Time steps used for the pyrazine, quartic oscillator, and
collinear NCO models were $0.5\,\text{a.u.}$, $10^{-3}$, and $2.5\,\text{a.u.%
}$, respectively. Also note that the branch of the square root in the
prefactor in Eq.~(\ref{eq:A_CDR}) was gradually adjusted in the course of
the propagation in order to ensure that the phase of the prefactor be
continuous in time. 

\section{Conclusions\label{sec:Final_remarks}}

We have introduced the CDRP, a rather accurate and efficient semiclassical
method for computing ultrafast time-resolved electronic spectra. The CDRP is
a two-stage refinement of the DR of fidelity amplitude: A prefactor
correction, which typically increases accuracy, is followed by a
cellularization procedure increasing efficiency (see Fig.~\ref{fig:CDRP
diagram}). The new method has the same computational cost per trajectory as
the two intermediate refinements, CDR and DRP; this cost is determined by
propagating the stability matrix and its derivatives. While the cost per
trajectory is significantly higher than the cost of each DR trajectory, the
reduction in the required number of trajectories can in many situations
result in higher efficiency compared with the DR.

The new methodology has been tested on several systems. In harmonic
potentials (Figs.~\ref{fig:SHO} and \ref{fig:SHO2D}), pyrazine-based model
(Fig.~\ref{fig:pyrazine}), and collinear NCO molecule (Fig.~\ref{fig:nco}),
the TRSE correlation functions and spectra computed with the CDRP were more
accurate and required fewer trajectories than the corresponding quantities
computed with the original DR. For harmonic potentials, analytical formulas
have been derived; particularly, we have shown that cellularized
calculations using a single trajectory are identical to the fully converged
noncellular methods since the second-order expansion of the DR phase is
exact. Moreover, in harmonic potentials the prefactor is the same for all
trajectories. In contrast, in systems with highly nonlinear or chaotic
dynamics, such as the quartic oscillator, the second-order approximation to
the semiclassical action $S_{\text{DR}}$ breaks down and its use can
decrease both the accuracy and efficiency. Interestingly, in such systems
the \textquotedblleft bare\textquotedblright\ DR can perform rather well
[see Fig.~\ref{fig:quartic}(a)], in agreement with previously published
results.\cite{Vanicek:2004,*Vanicek:2006}

An important result in its own right is the new simple, yet rigorous
cellularization scheme for the DR, in which the size and the sampling weight
of the Gaussian cells changes with their number. A similar cellularization
scheme using the inverse Weierstrass transform should be useful also for
more general quantum dynamics using semiclassical initial value
representations such as the Heller-Herman-Kluk-Kay propagator. 

\acknowledgments This research was supported by the Swiss NSF with Grant
No.~200021\_124936/1 and NCCR MUST (Molecular Ultrafast Science \&
Technology), and by the EPFL. We would like to thank M. Wehrle and T.
Zimmermann for discussions.%
\appendix

\section{Phase-space propagator\label{ap:PS propagator}}

Semiclassical propagator in position representation, known as the \emph{Van
Vleck propagator}, is given by the expression%
\begin{multline}
\hspace{-0.23cm}\langle q_{b}|e^{-i\hat{H}t/\hbar }|q_{a}\rangle _{\text{SC}%
}=\hspace{-0.2cm}\sum_{q_{a}\overset{j}{\leadsto }q_{b}}{(2i\pi \hbar
)^{-D/2}}\det \left( \frac{\partial ^{2}S_{j}}{\partial q_{a}\partial q_{b}}%
\right) ^{-1/2} \\
\times e^{iS_{j}(q_{a},q_{b};t)/\hbar -i\nu _{j}{\pi }/2},
\label{eq:Van Vleck propagator}
\end{multline}%
where the summation is performed over all trajectories $j$ of the classical
Hamiltonian $H$ starting from $q_{a}$ and arriving at $q_{b}$ after time $t$%
, $S_{j}$ is the classical action along the $j$th path, and $\nu _{j}$ is
its Morse index.

The \emph{phase-space propagator} is the Wigner transform of the evolution
operator, 
\begin{equation}
U_{W}(x,t)=\int d^{D}s\left\langle q-s/2\right\vert e^{-i\hat{H}t/\hbar
}\left\vert q+s/2\right\rangle e^{i\,s^{\mathsf{T}}\cdot p/\hbar }.
\label{eq:Wigner transform of U}
\end{equation}%
The integrand in the last equation includes the position propagator between $%
q+s/2$ to $q-s/2$. Using the Van Vleck propagator, we can obtain the
semiclassical expression for Eq.~(\ref{eq:Wigner transform of U}):\cite%
{Berry:1989,Almeida:1998} 
\begin{equation}
U_{\text{SC}}(\bar{x},t)=2^{D}\sum_{j}\left\vert \det \left(
I+M_{j}^{t}\right) \right\vert ^{-1/2}\,\exp \left[ \frac{i}{\hbar }S_{{%
\text{c}},j}(\bar{x},t)\right] ,  \label{eq:U semi}
\end{equation}%
where the sum runs over all paths $j$ centered at $\bar{x}$, i.e., paths for
which $(x^{0}+x^{t})/2=\bar{x}$ [see Fig.~\ref{fig:LE-areas U(x)}], $M^{t}$
is the \emph{stability matrix} of the flow $x^{0}\rightarrow x^{t}$, and the
function $S_{\text{c}}(\bar{x},t)$, called \emph{center-action}, is defined
as 
\begin{equation*}
S_{\text{c}}(\bar{x},t)=\oint p^{\mathsf{T}}\cdot
dq-\int_{0}^{t}H(x^{t^{\prime }},t^{\prime })\,dt^{\prime },
\end{equation*}%
where the first term is the symplectic area enclosed by a closed path
consisting of a trajectory centered at $\bar{x}$ and the chord connecting
this trajectory's final and initial points.

\begin{figure}[htbp]
\centering\includegraphics[width=0.9\columnwidth]{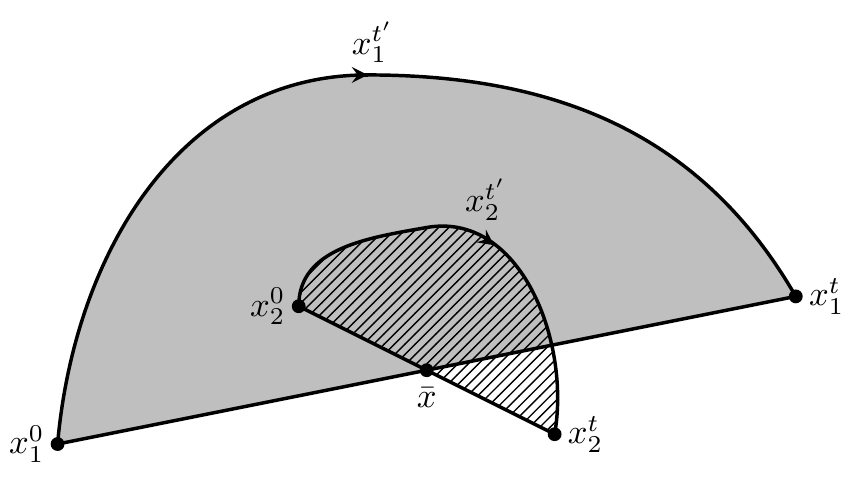} 
\caption{Geometrical interpretation of the semiclassical phase space
propagator.  Two trajectories ($x^{t'}_1$ and $x^{t'}_2$) contributing to $U_{\text{SC}}(\bar x,t)$ are shown;
 $\bar x$ is the midpoint of both. Geometrical parts of the center-actions $S_{{\text{c}},1}$ and $S_{{\text{c}},2}$ 
are displayed as filled and hatched areas, respectively.} \label%
{fig:LE-areas U(x)}
\end{figure}

In general, the center-action is multivalued and each of its branches is
associated with a classical trajectory centered at $\bar{x}$, as shown in
Fig.~\ref{fig:LE-areas U(x)}. The stability matrix, defined as $%
M^{t}:=\partial x^{t}/\partial x^{0}$, defines the local linearization of
the classical path in the tangent phase-space and the phase-space propagator
has caustics whenever $M^{t}$ has an eigenvalue $-1$.\cite{Berry:1989}
Moreover, $M^{t}$ is equal to the Cayley transform of $\frac{1}{2}J\cdot
\partial ^{2}S_{\text{c}}(\bar{x},t)/\partial \bar{x}^{2}$, 
\begin{equation}
M^{t}=\left( I-\frac{J}{2}\cdot \frac{\partial ^{2}S_{c}}{\partial {\bar{x}}%
^{2}}\right) \cdot \left( I+\frac{J}{2}\cdot \frac{\partial ^{2}S_{c}}{%
\partial {\bar{x}}^{2}}\right) ^{-1},
\end{equation}%
and the determinant in phase-space propagator (\ref{eq:U semi}) can be
written in terms of the center action as\cite{Almeida:1998} 
\begin{equation}
2^{2D}\left[ \det \left( I+M^{t}\right) \right] ^{-1}=\det \left( I+\frac{J}{%
2}\cdot \frac{\partial ^{2}S_{\text{c}}}{\partial \bar{x}^{2}}\right) .
\end{equation}%
The last relation follows from the fact that both $M^{t}$ and the Hessian of 
$S_{c}(\bar{x},t)$ define the same local linearization of the classical
equations of motion in a neighborhood of a classical trajectory $%
x^{t^{\prime }}$. This linearization is described by the mapping: 
\begin{equation}
x^{0}=\bar{x}+\frac{J}{2}\cdot \frac{\partial S_{\text{c}}}{\partial \bar{x}}%
\rightarrow x^{t}=\bar{x}-\frac{J}{2}\cdot \frac{\partial S_{\text{c}}}{%
\partial \bar{x}}.
\end{equation}

For short times, the Wigner transformation $\mathcal{E}_{W}(x,t)$ of the
echo operator can be approximated by a propagator~(\ref{eq:U semi}) with a
single classical trajectory,\cite{Almeida:1998,Zambrano_Almeida:2011} as in
Eq.~(\ref{eq:Echo_symbol}). 

\section{Derivatives of the DR phase in Eq.~(\protect\ref{eq:S_CDR})\label%
{ap:Derivatives}}

One of the main numerical prerequisites of both the DRP and CDR is the
second order expansion of the DR phase, $S_{\text{DR}}(x^{0},t)$, as
indicated in Eq.~(\ref{eq:S_CDR}). {Here we describe a symplectic numerical
procedure for obtaining the time derivatives of the phase-space derivatives $%
\partial ^{\lvert \alpha \rvert }S_{\text{DR}}(x^{0},t)/\partial {(x^{0})}%
^{\alpha }$ for $\lvert \alpha \rvert \leq 2$ (multi-index notation was
used). }

As in other semiclassical methods, the knowledge of the Hessian of the
potential is required for propagating the stability matrix $M^{t}$. Below we
show that in order to obtain the Hessian of $S_{\text{DR}}(x^{0},t)$ with
respect to $x^{0}$, third derivatives of the potential, $\nabla ^{3}V$, are
also needed. Although the third derivative is in principle required also in
Cellular Dynamics\cite{Heller:1991} and Cellularized Frozen Gaussian
approximation,\cite{Walton_Manolopoulos:1996} the associated computational
cost has led the authors of these methods to neglect the contribution of
terms depending on $\nabla ^{3}V$. However, as demonstrated in Fig.~\ref%
{fig:ds}(a), this contribution can be essential even in simple realistic
models such as the collinear NCO molecule.

First, consider components of the gradient of $S_{\text{DR}}$, 
\begin{equation}
\frac{\partial S_{\text{DR}}}{\partial x_{j}^{0}}=\int_{0}^{t}\!dt^{\prime
}\,\frac{\partial \Delta H}{\partial x_{k}^{t^{\prime }}}\frac{\partial
x_{k}^{t^{\prime }}}{\partial x_{j}^{0}}=-\int_{0}^{t}\!d{t^{\prime }}%
\,\Delta F_{k}^{t^{\prime }}M_{kj}^{t^{\prime }},  \label{eq:dS_gradient}
\end{equation}%
where $\Delta H=H_{2}-H_{1}$, $\Delta F^{t}\equiv -{\partial \Delta H}/{%
\partial {x}^{t}}$ is the force difference vector, and repeated indexes
imply summation. Similarly, the components of the Hessian of $S_{\text{DR}}$
are 
\begin{equation}
\frac{\partial ^{2}S_{\text{DR}}}{\partial x_{i}^{0}\partial x_{j}^{0}}%
=\int_{0}^{t}\!d{t^{\prime }}\left( \Delta {\mathcal{H}}_{ks}^{t^{\prime
}}M_{ki}^{t^{\prime }}M_{sj}^{t^{\prime }}-\Delta F_{k}^{t^{\prime
}}N_{k,ij}^{t^{\prime }}\right) ,  \label{eq:dS_hessian}
\end{equation}%
where $\Delta {\mathcal{H}}^{t}$ denotes the Hessian of $\Delta H$ at time $%
t $ and 
\begin{equation}
N_{k,ij}^{t}:=\frac{\partial ^{2}x_{k}^{t}}{\partial x_{i}^{0}\partial
x_{j}^{0}}=\frac{\partial }{\partial x_{i}^{0}}M_{kj}^{t}.
\end{equation}%
While the {time integrals in Eqs.~(\ref{eq:dS_gradient}) and (\ref%
{eq:dS_hessian}) are evaluated using composite Newton-Cotes formulas, the
integrands can be propagated symplectically. The algorithm for }$N^{t}$
propagation, e.g., is obtained by applying the chain rule to the preceding
equation, 
\begin{equation}
N_{k,ij}^{t+\delta t}=\frac{\partial x_{k}^{t+\delta t}}{\partial x_{s}^{t}}%
N_{s,ij}^{t}+\frac{\partial ^{2}x_{k}^{t+\delta t}}{\partial
x_{n}^{t}\partial x_{s}^{t}}M_{ni}^{t}M_{sj}^{t},  \label{eq:N}
\end{equation}%
whereas the symplectic propagation scheme for the stability matrix was
described previously:\cite{Brewer_Hulme:1997,Sulc_Vanicek:2012} 
\begin{equation}
M_{ij}^{t+\delta t}=\frac{\partial x_{i}^{t+\delta t}}{\partial x_{k}^{t}}%
M_{kj}^{t}.  \label{eq:M}
\end{equation}%
Derivatives of phase-space coordinates in Eqs.~(\ref{eq:N}) and (\ref{eq:M})
are obtained from symplectic integrators for $q$ and $p$, which are for
standard Hamiltonians of the form $\sum_{i}p_{i}^{2}/2m_{i}+V(q)$ based on a
Lie-Trotter-type\cite{Trotter:1959} decomposition of a short-time propagator
into elementary steps within which the system is propagated under the
influence of either the kinetic or the potential term only. Action of the
kinetic term $\sum_{i}p_{i}^{2}/2m_{i}$ for time $\delta t$ results in a
phase-space shear preserving the momentum, 
\begin{equation}
(q^{t+\delta t}\!,p^{t+\delta t})=\left( q^{t}+m^{-1}\!\cdot p^{t}\,\delta
{}t,p^{t}\right) ,  \label{eq:P integrator}
\end{equation}%
whereas the action of the potential term $V(q)$ changes momentum and
preserves position: 
\begin{equation}
(q^{t+\delta t}\!,p^{t+\delta t})=\left( q^{t}\!,p^{t}-\frac{\partial
V(q^{t})}{\partial q^{t}}\delta t\right) .  \label{eq:P_prop}
\end{equation}%
Since the only nonlinear dependence of $(q^{t+\delta t},\,p^{t+\delta t})$
on $(q^{t},\,p^{t})$ stems from the presence of the potential gradient in
Eq.~(\ref{eq:P_prop}), the second derivative terms in Eq.~(\ref{eq:N}) are
nonzero only during the \textquotedblleft p-propagation\textquotedblright\ (%
\ref{eq:P_prop}) and explicitly involve derivatives of the Hessian: 
\begin{equation}
\frac{\partial ^{2}p_{k}^{t+\delta t}}{\partial q_{i}^{t}\partial q_{j}^{t}}%
=-\delta t\frac{\partial ^{3}V(q^{t})}{\partial q_{k}^{t}\partial
q_{i}^{t}\partial q_{j}^{t}}.
\end{equation}%
As already mentioned, these third derivatives of the potential, which should
appear in other semiclassical propagation schemes\cite%
{Heller:1991,Walton_Manolopoulos:1996} as well, are usually neglected in
order to reduce computational cost. Yet, in Section \ref{subsec:nco} we have
shown that they can play an essential role even in rather simple systems
such as the NCO.

\section{\label{ap:Proof}Complex conjugate of Eq.~(\protect\ref{eq:DRP
formula})}

As discussed in Subsec.~\ref{subsec:DRP}, switching the roles of the PESs in
Eq.~(\ref{eq:f(t,tau)}) for the correlation function corresponds (for $\tau
=0$) to taking the complex conjugate of this equation. Likewise, when one
interchanges the PESs, the DR phase and hence the matrix $B_{x^{0}}^{t}$
change the sign. However, since the prefactor $A_{\text{DRP}}$ in Eq.~(\ref%
{eq:DRP formula}) is real, it might seem that the DRP is incompatible with
this operation.

Here we demonstrate that this is not the case by proving that $\det
(I+J\cdot B_{x^{0}}^{t})=\det (I-J\cdot B_{x^{0}}^{t})$. To this end,
consider a~general, symmetric, $2D\times 2D$ matrix $A$ and let $a$ denote
any of the eigenvalues of $J\cdot A$. Then 
\begin{align}
0& =\det (J\cdot A-aI)=\det (A-aJ^{\mathsf{T}})  \notag
\label{eq:app_B_proof} \\
& =\det (A+aJ)=\det (A+aJ)^{\mathsf{T}}  \notag \\
& =\det (A+aJ^{\mathsf{T}})=\det (J\cdot A+aI),
\end{align}%
where we have used the properties $-J=J^{\mathsf{T}}=J^{-1}$, $\det {J}=1$,
and that taking the transpose of an arbitrary square matrix does not affect
its determinant.

Equation~(\ref{eq:app_B_proof}) shows that the eigenvalues of $J\cdot A$
come in pairs $(a,-a)$. This directly implies that $\det (I+J\cdot A)=\det
(I-J\cdot A)$. [This also follows from setting $a=1$ in Eq.~(\ref%
{eq:app_B_proof}) and using the fact that the matrix $J\cdot A$ is of even
order.] 


%


\begin{thebibliography}{75}%
\makeatletter
\providecommand \@ifxundefined [1]{%
 \@ifx{#1\undefined}
}%
\providecommand \@ifnum [1]{%
 \ifnum #1\expandafter \@firstoftwo
 \else \expandafter \@secondoftwo
 \fi
}%
\providecommand \@ifx [1]{%
 \ifx #1\expandafter \@firstoftwo
 \else \expandafter \@secondoftwo
 \fi
}%
\providecommand \natexlab [1]{#1}%
\providecommand \enquote  [1]{``#1''}%
\providecommand \bibnamefont  [1]{#1}%
\providecommand \bibfnamefont [1]{#1}%
\providecommand \citenamefont [1]{#1}%
\providecommand \href@noop [0]{\@secondoftwo}%
\providecommand \href [0]{\begingroup \@sanitize@url \@href}%
\providecommand \@href[1]{\@@startlink{#1}\@@href}%
\providecommand \@@href[1]{\endgroup#1\@@endlink}%
\providecommand \@sanitize@url [0]{\catcode `\\12\catcode `\$12\catcode
  `\&12\catcode `\#12\catcode `\^12\catcode `\_12\catcode `\%12\relax}%
\providecommand \@@startlink[1]{}%
\providecommand \@@endlink[0]{}%
\providecommand \url  [0]{\begingroup\@sanitize@url \@url }%
\providecommand \@url [1]{\endgroup\@href {#1}{\urlprefix }}%
\providecommand \urlprefix  [0]{URL }%
\providecommand \Eprint [0]{\href }%
\providecommand \doibase [0]{http://dx.doi.org/}%
\providecommand \selectlanguage [0]{\@gobble}%
\providecommand \bibinfo  [0]{\@secondoftwo}%
\providecommand \bibfield  [0]{\@secondoftwo}%
\providecommand \translation [1]{[#1]}%
\providecommand \BibitemOpen [0]{}%
\providecommand \bibitemStop [0]{}%
\providecommand \bibitemNoStop [0]{.\EOS\space}%
\providecommand \EOS [0]{\spacefactor3000\relax}%
\providecommand \BibitemShut  [1]{\csname bibitem#1\endcsname}%
\let\auto@bib@innerbib\@empty
\bibitem [{\citenamefont {Bisgaard}\ \emph {et~al.}(2009)\citenamefont
  {Bisgaard}, \citenamefont {Clarkin}, \citenamefont {Wu}, \citenamefont {Lee},
  \citenamefont {Gessner}, \citenamefont {Hayden},\ and\ \citenamefont
  {Stolow}}]{Bisgaard_Clarkin:2009}%
  \BibitemOpen
  \bibfield  {author} {\bibinfo {author} {\bibfnamefont {C.~Z.}\ \bibnamefont
  {Bisgaard}}, \bibinfo {author} {\bibfnamefont {O.~J.}\ \bibnamefont
  {Clarkin}}, \bibinfo {author} {\bibfnamefont {G.}~\bibnamefont {Wu}},
  \bibinfo {author} {\bibfnamefont {A.~M.~D.}\ \bibnamefont {Lee}}, \bibinfo
  {author} {\bibfnamefont {O.}~\bibnamefont {Gessner}}, \bibinfo {author}
  {\bibfnamefont {C.~C.}\ \bibnamefont {Hayden}}, \ and\ \bibinfo {author}
  {\bibfnamefont {A.}~\bibnamefont {Stolow}},\ }\href {\doibase
  10.1126/science.1169183} {\bibfield  {journal} {\bibinfo  {journal}
  {Science}\ }\textbf {\bibinfo {volume} {323}},\ \bibinfo {pages} {1464}
  (\bibinfo {year} {2009})}\BibitemShut {NoStop}%
\bibitem [{\citenamefont {Bressler}\ \emph {et~al.}(2009)\citenamefont
  {Bressler}, \citenamefont {Milne}, \citenamefont {Pham}, \citenamefont
  {ElNahhas}, \citenamefont {van~der Veen}, \citenamefont {Gawelda},
  \citenamefont {Johnson}, \citenamefont {Beaud}, \citenamefont {Grolimund},
  \citenamefont {Kaiser}, \citenamefont {Borca}, \citenamefont {Ingold},
  \citenamefont {Abela},\ and\ \citenamefont {Chergui}}]{Bressler_Milne:2009}%
  \BibitemOpen
  \bibfield  {author} {\bibinfo {author} {\bibfnamefont {C.}~\bibnamefont
  {Bressler}}, \bibinfo {author} {\bibfnamefont {C.}~\bibnamefont {Milne}},
  \bibinfo {author} {\bibfnamefont {V.-T.}\ \bibnamefont {Pham}}, \bibinfo
  {author} {\bibfnamefont {A.}~\bibnamefont {ElNahhas}}, \bibinfo {author}
  {\bibfnamefont {R.~M.}\ \bibnamefont {van~der Veen}}, \bibinfo {author}
  {\bibfnamefont {W.}~\bibnamefont {Gawelda}}, \bibinfo {author} {\bibfnamefont
  {S.}~\bibnamefont {Johnson}}, \bibinfo {author} {\bibfnamefont
  {P.}~\bibnamefont {Beaud}}, \bibinfo {author} {\bibfnamefont
  {D.}~\bibnamefont {Grolimund}}, \bibinfo {author} {\bibfnamefont
  {M.}~\bibnamefont {Kaiser}}, \bibinfo {author} {\bibfnamefont {C.~N.}\
  \bibnamefont {Borca}}, \bibinfo {author} {\bibfnamefont {G.}~\bibnamefont
  {Ingold}}, \bibinfo {author} {\bibfnamefont {R.}~\bibnamefont {Abela}}, \
  and\ \bibinfo {author} {\bibfnamefont {M.}~\bibnamefont {Chergui}},\ }\href
  {\doibase 10.1126/science.1165733} {\bibfield  {journal} {\bibinfo  {journal}
  {Science}\ }\textbf {\bibinfo {volume} {323}},\ \bibinfo {pages} {489}
  (\bibinfo {year} {2009})}\BibitemShut {NoStop}%
\bibitem [{\citenamefont {Carbone}, \citenamefont {Kwon},\ and\ \citenamefont
  {Zewail}(2009)}]{Carbone_Kwon:2009}%
  \BibitemOpen
  \bibfield  {author} {\bibinfo {author} {\bibfnamefont {F.}~\bibnamefont
  {Carbone}}, \bibinfo {author} {\bibfnamefont {O.-H.}\ \bibnamefont {Kwon}}, \
  and\ \bibinfo {author} {\bibfnamefont {A.~H.}\ \bibnamefont {Zewail}},\
  }\href {\doibase 10.1126/science.1175005} {\bibfield  {journal} {\bibinfo
  {journal} {Science}\ }\textbf {\bibinfo {volume} {325}},\ \bibinfo {pages}
  {181} (\bibinfo {year} {2009})}\BibitemShut {NoStop}%
\bibitem [{\citenamefont {Miller}(1970)}]{Miller:1970}%
  \BibitemOpen
  \bibfield  {author} {\bibinfo {author} {\bibfnamefont {W.~H.}\ \bibnamefont
  {Miller}},\ }\href {\doibase 10.1063/1.1674535} {\bibfield  {journal}
  {\bibinfo  {journal} {J.~Chem.\ Phys.}\ }\textbf {\bibinfo {volume} {53}},\
  \bibinfo {pages} {3578} (\bibinfo {year} {1970})}\BibitemShut {NoStop}%
\bibitem [{\citenamefont {Heller}(1981)}]{Heller:1981}%
  \BibitemOpen
  \bibfield  {author} {\bibinfo {author} {\bibfnamefont {E.~J.}\ \bibnamefont
  {Heller}},\ }\href {\doibase 10.1063/1.442382} {\bibfield  {journal}
  {\bibinfo  {journal} {J.~Chem.\ Phys.}\ }\textbf {\bibinfo {volume} {75}},\
  \bibinfo {pages} {2923} (\bibinfo {year} {1981})}\BibitemShut {NoStop}%
\bibitem [{\citenamefont {Herman}\ and\ \citenamefont
  {Kluk}(1984)}]{Herman_Kluk:1984}%
  \BibitemOpen
  \bibfield  {author} {\bibinfo {author} {\bibfnamefont {M.~F.}\ \bibnamefont
  {Herman}}\ and\ \bibinfo {author} {\bibfnamefont {E.}~\bibnamefont {Kluk}},\
  }\href {\doibase 10.1016/0301-0104(84)80039-7} {\bibfield  {journal}
  {\bibinfo  {journal} {Chem.\ Phys.}\ }\textbf {\bibinfo {volume} {91}},\
  \bibinfo {pages} {27} (\bibinfo {year} {1984})}\BibitemShut {NoStop}%
\bibitem [{\citenamefont {Miller}(2001)}]{Miller:2001}%
  \BibitemOpen
  \bibfield  {author} {\bibinfo {author} {\bibfnamefont {W.~H.}\ \bibnamefont
  {Miller}},\ }\href@noop {} {\bibfield  {journal} {\bibinfo  {journal}
  {J.~Phys.\ Chem.}\ }\textbf {\bibinfo {volume} {105}},\ \bibinfo {pages}
  {2942} (\bibinfo {year} {2001})}\BibitemShut {NoStop}%
\bibitem [{\citenamefont {Herman}(1994)}]{Herman:1994}%
  \BibitemOpen
  \bibfield  {author} {\bibinfo {author} {\bibfnamefont {M.~F.}\ \bibnamefont
  {Herman}},\ }\href {\doibase 10.1146/annurev.pc.45.100194.000503} {\bibfield
  {journal} {\bibinfo  {journal} {Annu.\ Rev.\ Phys.\ Chem.}\ }\textbf
  {\bibinfo {volume} {45}},\ \bibinfo {pages} {83} (\bibinfo {year}
  {1994})}\BibitemShut {NoStop}%
\bibitem [{\citenamefont {Thoss}\ and\ \citenamefont
  {Wang}(2004)}]{Thoss_Wang:2004}%
  \BibitemOpen
  \bibfield  {author} {\bibinfo {author} {\bibfnamefont {M.}~\bibnamefont
  {Thoss}}\ and\ \bibinfo {author} {\bibfnamefont {H.}~\bibnamefont {Wang}},\
  }\href {\doibase 10.1146/annurev.physchem.55.091602.094429} {\bibfield
  {journal} {\bibinfo  {journal} {Annu.\ Rev.\ Phys.\ Chem.}\ }\textbf
  {\bibinfo {volume} {55}},\ \bibinfo {pages} {299} (\bibinfo {year}
  {2004})}\BibitemShut {NoStop}%
\bibitem [{\citenamefont {Kay}(2005)}]{Kay:2005}%
  \BibitemOpen
  \bibfield  {author} {\bibinfo {author} {\bibfnamefont {K.~G.}\ \bibnamefont
  {Kay}},\ }\href {\doibase 10.1146/annurev.physchem.56.092503.141257}
  {\bibfield  {journal} {\bibinfo  {journal} {Annu.\ Rev.\ Phys.\ Chem.}\
  }\textbf {\bibinfo {volume} {56}},\ \bibinfo {pages} {255} (\bibinfo {year}
  {2005})}\BibitemShut {NoStop}%
\bibitem [{\citenamefont {Ceotto}, \citenamefont {Zhuang},\ and\ \citenamefont
  {Hase}(2013)}]{Ceotto_Zhuang:2013}%
  \BibitemOpen
  \bibfield  {author} {\bibinfo {author} {\bibfnamefont {M.}~\bibnamefont
  {Ceotto}}, \bibinfo {author} {\bibfnamefont {Y.}~\bibnamefont {Zhuang}}, \
  and\ \bibinfo {author} {\bibfnamefont {W.~L.}\ \bibnamefont {Hase}},\ }\href
  {\doibase 10.1063/1.4789759} {\bibfield  {journal} {\bibinfo  {journal}
  {J.~Chem.\ Phys.}\ }\textbf {\bibinfo {volume} {138}},\ \bibinfo {eid}
  {054116} (\bibinfo {year} {2013})}\BibitemShut {NoStop}%
\bibitem [{\citenamefont {Van\'{\i}\v{c}ek}(2004)}]{Vanicek:2004}%
  \BibitemOpen
  \bibfield  {author} {\bibinfo {author} {\bibfnamefont {J.}~\bibnamefont
  {Van\'{\i}\v{c}ek}},\ }\href {\doibase 10.1103/PhysRevE.70.055201} {\bibfield
   {journal} {\bibinfo  {journal} {Phys.\ Rev.~E}\ }\textbf {\bibinfo {volume}
  {70}},\ \bibinfo {pages} {055201} (\bibinfo {year} {2004})}\BibitemShut
  {NoStop}%
\bibitem [{\citenamefont {Van\'{\i}\v{c}ek}(2006)}]{Vanicek:2006}%
  \BibitemOpen
  \bibfield  {author} {\bibinfo {author} {\bibfnamefont {J.}~\bibnamefont
  {Van\'{\i}\v{c}ek}},\ }\href {\doibase 10.1103/PhysRevE.73.046204} {\bibfield
   {journal} {\bibinfo  {journal} {Phys.\ Rev.~E}\ }\textbf {\bibinfo {volume}
  {73}},\ \bibinfo {pages} {046204} (\bibinfo {year} {2006})}\BibitemShut
  {NoStop}%
\bibitem [{\citenamefont {Wehrle}, \citenamefont {\v{S}ulc},\ and\
  \citenamefont {Van\'{\i}\v{c}ek}(2011)}]{Wehrle_Sulc:2011}%
  \BibitemOpen
  \bibfield  {author} {\bibinfo {author} {\bibfnamefont {M.}~\bibnamefont
  {Wehrle}}, \bibinfo {author} {\bibfnamefont {M.}~\bibnamefont {\v{S}ulc}}, \
  and\ \bibinfo {author} {\bibfnamefont {J.}~\bibnamefont {Van\'{\i}\v{c}ek}},\
  }\href {\doibase 10.2533/chimia.2011.334} {\bibfield  {journal} {\bibinfo
  {journal} {Chimia}\ }\textbf {\bibinfo {volume} {65}},\ \bibinfo {pages}
  {334} (\bibinfo {year} {2011})}\BibitemShut {NoStop}%
\bibitem [{\citenamefont {\v{S}ulc}\ and\ \citenamefont
  {Van\'{\i}\v{c}ek}(2012)}]{Sulc_Vanicek:2012}%
  \BibitemOpen
  \bibfield  {author} {\bibinfo {author} {\bibfnamefont {M.}~\bibnamefont
  {\v{S}ulc}}\ and\ \bibinfo {author} {\bibfnamefont {J.}~\bibnamefont
  {Van\'{\i}\v{c}ek}},\ }\href {\doibase 10.1080/00268976.2012.668971}
  {\bibfield  {journal} {\bibinfo  {journal} {Mol.\ Phys.}\ }\textbf {\bibinfo
  {volume} {110}},\ \bibinfo {pages} {945} (\bibinfo {year}
  {2012})}\BibitemShut {NoStop}%
\bibitem [{\citenamefont {Van\'{\i}\v{c}ek}\ and\ \citenamefont
  {Heller}(2003)}]{Vanicek_Heller:2003}%
  \BibitemOpen
  \bibfield  {author} {\bibinfo {author} {\bibfnamefont {J.}~\bibnamefont
  {Van\'{\i}\v{c}ek}}\ and\ \bibinfo {author} {\bibfnamefont {E.~J.}\
  \bibnamefont {Heller}},\ }\href {\doibase 10.1103/PhysRevE.68.056208}
  {\bibfield  {journal} {\bibinfo  {journal} {Phys.\ Rev.~E}\ }\textbf
  {\bibinfo {volume} {68}},\ \bibinfo {pages} {056208} (\bibinfo {year}
  {2003})}\BibitemShut {NoStop}%
\bibitem [{\citenamefont {Miller}\ and\ \citenamefont
  {Smith}(1978)}]{Miller_Smith:1978}%
  \BibitemOpen
  \bibfield  {author} {\bibinfo {author} {\bibfnamefont {W.~H.}\ \bibnamefont
  {Miller}}\ and\ \bibinfo {author} {\bibfnamefont {F.~T.}\ \bibnamefont
  {Smith}},\ }\href {\doibase 10.1103/PhysRevA.17.939} {\bibfield  {journal}
  {\bibinfo  {journal} {Phys.\ Rev.~A}\ }\textbf {\bibinfo {volume} {17}},\
  \bibinfo {pages} {939} (\bibinfo {year} {1978})}\BibitemShut {NoStop}%
\bibitem [{\citenamefont {Hubbard}\ and\ \citenamefont
  {Miller}(1983)}]{Hubbard_Miller:1983}%
  \BibitemOpen
  \bibfield  {author} {\bibinfo {author} {\bibfnamefont {L.~M.}\ \bibnamefont
  {Hubbard}}\ and\ \bibinfo {author} {\bibfnamefont {W.~H.}\ \bibnamefont
  {Miller}},\ }\href {\doibase 10.1063/1.444976} {\bibfield  {journal}
  {\bibinfo  {journal} {J.~Chem.\ Phys.}\ }\textbf {\bibinfo {volume} {78}},\
  \bibinfo {pages} {1801} (\bibinfo {year} {1983})}\BibitemShut {NoStop}%
\bibitem [{\citenamefont {Mukamel}(1982)}]{Mukamel:1982}%
  \BibitemOpen
  \bibfield  {author} {\bibinfo {author} {\bibfnamefont {S.}~\bibnamefont
  {Mukamel}},\ }\href {\doibase 10.1063/1.443638} {\bibfield  {journal}
  {\bibinfo  {journal} {J.~Chem.\ Phys.}\ }\textbf {\bibinfo {volume} {77}},\
  \bibinfo {pages} {173} (\bibinfo {year} {1982})}\BibitemShut {NoStop}%
\bibitem [{\citenamefont {Mukamel}(1999)}]{book_Mukamel}%
  \BibitemOpen
  \bibfield  {author} {\bibinfo {author} {\bibfnamefont {S.}~\bibnamefont
  {Mukamel}},\ }\href@noop {} {\emph {\bibinfo {title} {Principles of nonlinear
  optical spectroscopy}}},\ \bibinfo {edition} {1st}\ ed.\ (\bibinfo
  {publisher} {Oxford University Press},\ \bibinfo {address} {New York},\
  \bibinfo {year} {1999})\BibitemShut {NoStop}%
\bibitem [{\citenamefont {Shemetulskis}\ and\ \citenamefont
  {Loring}(1992)}]{Shemetulskis_Loring:1992}%
  \BibitemOpen
  \bibfield  {author} {\bibinfo {author} {\bibfnamefont {N.~E.}\ \bibnamefont
  {Shemetulskis}}\ and\ \bibinfo {author} {\bibfnamefont {R.~F.}\ \bibnamefont
  {Loring}},\ }\href {\doibase 10.1063/1.463248} {\bibfield  {journal}
  {\bibinfo  {journal} {J.~Chem.\ Phys.}\ }\textbf {\bibinfo {volume} {97}},\
  \bibinfo {pages} {1217} (\bibinfo {year} {1992})}\BibitemShut {NoStop}%
\bibitem [{\citenamefont {Rost}(1995)}]{Rost:1995}%
  \BibitemOpen
  \bibfield  {author} {\bibinfo {author} {\bibfnamefont {J.~M.}\ \bibnamefont
  {Rost}},\ }\href {http://stacks.iop.org/0953-4075/28/i=19/a=002} {\bibfield
  {journal} {\bibinfo  {journal} {J.~Phys.~B}\ }\textbf {\bibinfo {volume}
  {28}},\ \bibinfo {pages} {L601} (\bibinfo {year} {1995})}\BibitemShut
  {NoStop}%
\bibitem [{\citenamefont {Wang}, \citenamefont {Sun},\ and\ \citenamefont
  {Miller}(1998)}]{Wang_Sun:1998}%
  \BibitemOpen
  \bibfield  {author} {\bibinfo {author} {\bibfnamefont {H.}~\bibnamefont
  {Wang}}, \bibinfo {author} {\bibfnamefont {X.}~\bibnamefont {Sun}}, \ and\
  \bibinfo {author} {\bibfnamefont {W.~H.}\ \bibnamefont {Miller}},\ }\href
  {\doibase 10.1063/1.476447} {\bibfield  {journal} {\bibinfo  {journal}
  {J.~Chem.\ Phys.}\ }\textbf {\bibinfo {volume} {108}},\ \bibinfo {pages}
  {9726} (\bibinfo {year} {1998})}\BibitemShut {NoStop}%
\bibitem [{\citenamefont {Li}, \citenamefont {Fang},\ and\ \citenamefont
  {Martens}(1996)}]{Li_Fang:1996}%
  \BibitemOpen
  \bibfield  {author} {\bibinfo {author} {\bibfnamefont {Z.}~\bibnamefont
  {Li}}, \bibinfo {author} {\bibfnamefont {J.-Y.}\ \bibnamefont {Fang}}, \ and\
  \bibinfo {author} {\bibfnamefont {C.~C.}\ \bibnamefont {Martens}},\ }\href
  {\doibase 10.1063/1.471407} {\bibfield  {journal} {\bibinfo  {journal}
  {J.~Chem.\ Phys.}\ }\textbf {\bibinfo {volume} {104}},\ \bibinfo {pages}
  {6919} (\bibinfo {year} {1996})}\BibitemShut {NoStop}%
\bibitem [{\citenamefont {Egorov}, \citenamefont {Rabani},\ and\ \citenamefont
  {Berne}(1998)}]{Egorov_Rabani:1998}%
  \BibitemOpen
  \bibfield  {author} {\bibinfo {author} {\bibfnamefont {S.~A.}\ \bibnamefont
  {Egorov}}, \bibinfo {author} {\bibfnamefont {E.}~\bibnamefont {Rabani}}, \
  and\ \bibinfo {author} {\bibfnamefont {B.~J.}\ \bibnamefont {Berne}},\ }\href
  {\doibase 10.1063/1.475512} {\bibfield  {journal} {\bibinfo  {journal}
  {J.~Chem.\ Phys.}\ }\textbf {\bibinfo {volume} {108}},\ \bibinfo {pages}
  {1407} (\bibinfo {year} {1998})}\BibitemShut {NoStop}%
\bibitem [{\citenamefont {Egorov}, \citenamefont {Rabani},\ and\ \citenamefont
  {Berne}(1999)}]{Egorov_Rabani:1999}%
  \BibitemOpen
  \bibfield  {author} {\bibinfo {author} {\bibfnamefont {S.~A.}\ \bibnamefont
  {Egorov}}, \bibinfo {author} {\bibfnamefont {E.}~\bibnamefont {Rabani}}, \
  and\ \bibinfo {author} {\bibfnamefont {B.~J.}\ \bibnamefont {Berne}},\ }\href
  {\doibase 10.1063/1.478420} {\bibfield  {journal} {\bibinfo  {journal}
  {J.~Chem.\ Phys.}\ }\textbf {\bibinfo {volume} {110}},\ \bibinfo {pages}
  {5238} (\bibinfo {year} {1999})}\BibitemShut {NoStop}%
\bibitem [{\citenamefont {Shi}\ and\ \citenamefont
  {Geva}(2005)}]{Shi_Geva:2005}%
  \BibitemOpen
  \bibfield  {author} {\bibinfo {author} {\bibfnamefont {Q.}~\bibnamefont
  {Shi}}\ and\ \bibinfo {author} {\bibfnamefont {E.}~\bibnamefont {Geva}},\
  }\href {\doibase 10.1063/1.1843813} {\bibfield  {journal} {\bibinfo
  {journal} {J.~Chem.\ Phys.}\ }\textbf {\bibinfo {volume} {122}},\ \bibinfo
  {eid} {064506} (\bibinfo {year} {2005})}\BibitemShut {NoStop}%
\bibitem [{\citenamefont {Poulsen}, \citenamefont {Nyman},\ and\ \citenamefont
  {Rossky}(2003)}]{Poulsen_Nyman:2003}%
  \BibitemOpen
  \bibfield  {author} {\bibinfo {author} {\bibfnamefont {J.~A.}\ \bibnamefont
  {Poulsen}}, \bibinfo {author} {\bibfnamefont {G.}~\bibnamefont {Nyman}}, \
  and\ \bibinfo {author} {\bibfnamefont {P.~J.}\ \bibnamefont {Rossky}},\
  }\href {\doibase 10.1063/1.1626631} {\bibfield  {journal} {\bibinfo
  {journal} {J.~Chem.\ Phys.}\ }\textbf {\bibinfo {volume} {119}},\ \bibinfo
  {pages} {12179} (\bibinfo {year} {2003})}\BibitemShut {NoStop}%
\bibitem [{\citenamefont {Bonella}\ and\ \citenamefont
  {Coker}(2005)}]{Bonella_Coker:2005}%
  \BibitemOpen
  \bibfield  {author} {\bibinfo {author} {\bibfnamefont {S.}~\bibnamefont
  {Bonella}}\ and\ \bibinfo {author} {\bibfnamefont {D.~F.}\ \bibnamefont
  {Coker}},\ }\href {\doibase 10.1063/1.1896948} {\bibfield  {journal}
  {\bibinfo  {journal} {J.~Chem.\ Phys.}\ }\textbf {\bibinfo {volume} {122}},\
  \bibinfo {eid} {194102} (\bibinfo {year} {2005})}\BibitemShut {NoStop}%
\bibitem [{\citenamefont {Zimmermann}\ and\ \citenamefont
  {Van\'{\i}\v{c}ek}(2012{\natexlab{a}})}]{Zimmermann_Vanicek:2012}%
  \BibitemOpen
  \bibfield  {author} {\bibinfo {author} {\bibfnamefont {T.}~\bibnamefont
  {Zimmermann}}\ and\ \bibinfo {author} {\bibfnamefont {J.}~\bibnamefont
  {Van\'{\i}\v{c}ek}},\ }\href {\doibase 10.1063/1.3690458} {\bibfield
  {journal} {\bibinfo  {journal} {J.~Chem.\ Phys.}\ }\textbf {\bibinfo {volume}
  {136}},\ \bibinfo {eid} {094106} (\bibinfo {year}
  {2012}{\natexlab{a}})}\BibitemShut {NoStop}%
\bibitem [{\citenamefont {Zimmermann}\ and\ \citenamefont
  {Van\'{\i}\v{c}ek}(2012{\natexlab{b}})}]{Zimmermann_Vanicek:2012a}%
  \BibitemOpen
  \bibfield  {author} {\bibinfo {author} {\bibfnamefont {T.}~\bibnamefont
  {Zimmermann}}\ and\ \bibinfo {author} {\bibfnamefont {J.}~\bibnamefont
  {Van\'{\i}\v{c}ek}},\ }\href {\doibase 10.1063/1.4738878} {\bibfield
  {journal} {\bibinfo  {journal} {J.~Chem.\ Phys.}\ }\textbf {\bibinfo {volume}
  {137}},\ \bibinfo {eid} {22A516} (\bibinfo {year}
  {2012}{\natexlab{b}})}\BibitemShut {NoStop}%
\bibitem [{\citenamefont {Wang}\ \emph {et~al.}(2005)\citenamefont {Wang},
  \citenamefont {Casati}, \citenamefont {Li},\ and\ \citenamefont
  {Prosen}}]{Wang_Casati:2005}%
  \BibitemOpen
  \bibfield  {author} {\bibinfo {author} {\bibfnamefont {W.}~\bibnamefont
  {Wang}}, \bibinfo {author} {\bibfnamefont {G.}~\bibnamefont {Casati}},
  \bibinfo {author} {\bibfnamefont {B.}~\bibnamefont {Li}}, \ and\ \bibinfo
  {author} {\bibfnamefont {T.}~\bibnamefont {Prosen}},\ }\href {\doibase
  10.1103/PhysRevE.71.037202} {\bibfield  {journal} {\bibinfo  {journal}
  {Phys.\ Rev.~E}\ }\textbf {\bibinfo {volume} {71}},\ \bibinfo {pages}
  {037202} (\bibinfo {year} {2005})}\BibitemShut {NoStop}%
\bibitem [{\citenamefont {Ares}\ and\ \citenamefont
  {Wisniacki}(2009)}]{Ares_Wisniacki:2009}%
  \BibitemOpen
  \bibfield  {author} {\bibinfo {author} {\bibfnamefont {N.}~\bibnamefont
  {Ares}}\ and\ \bibinfo {author} {\bibfnamefont {D.~A.}\ \bibnamefont
  {Wisniacki}},\ }\href {\doibase 10.1103/PhysRevE.80.046216} {\bibfield
  {journal} {\bibinfo  {journal} {Phys.\ Rev.~E}\ }\textbf {\bibinfo {volume}
  {80}},\ \bibinfo {pages} {046216} (\bibinfo {year} {2009})}\BibitemShut
  {NoStop}%
\bibitem [{\citenamefont {Wisniacki}, \citenamefont {Ares},\ and\ \citenamefont
  {Vergini}(2010)}]{Wisniacki_Ares:2010}%
  \BibitemOpen
  \bibfield  {author} {\bibinfo {author} {\bibfnamefont {D.~A.}\ \bibnamefont
  {Wisniacki}}, \bibinfo {author} {\bibfnamefont {N.}~\bibnamefont {Ares}}, \
  and\ \bibinfo {author} {\bibfnamefont {E.~G.}\ \bibnamefont {Vergini}},\
  }\href {\doibase 10.1103/PhysRevLett.104.254101} {\bibfield  {journal}
  {\bibinfo  {journal} {Phys.\ Rev.\ Lett.}\ }\textbf {\bibinfo {volume}
  {104}},\ \bibinfo {pages} {254101} (\bibinfo {year} {2010})}\BibitemShut
  {NoStop}%
\bibitem [{\citenamefont {Garc\'{i}a-Mata}\ and\ \citenamefont
  {Wisniacki}(2011)}]{Garcia-Mata_Wisniacki:2011}%
  \BibitemOpen
  \bibfield  {author} {\bibinfo {author} {\bibfnamefont {I.}~\bibnamefont
  {Garc\'{i}a-Mata}}\ and\ \bibinfo {author} {\bibfnamefont {D.~A.}\
  \bibnamefont {Wisniacki}},\ }\href
  {http://stacks.iop.org/1751-8121/44/i=31/a=315101} {\bibfield  {journal}
  {\bibinfo  {journal} {J.~Phys.~A}\ }\textbf {\bibinfo {volume} {44}},\
  \bibinfo {pages} {315101} (\bibinfo {year} {2011})}\BibitemShut {NoStop}%
\bibitem [{\citenamefont {Mollica}\ and\ \citenamefont
  {Van\'{\i}\v{c}ek}(2011)}]{Mollica_Vanicek:2011}%
  \BibitemOpen
  \bibfield  {author} {\bibinfo {author} {\bibfnamefont {C.}~\bibnamefont
  {Mollica}}\ and\ \bibinfo {author} {\bibfnamefont {J.}~\bibnamefont
  {Van\'{\i}\v{c}ek}},\ }\href {\doibase 10.1103/PhysRevLett.107.214101}
  {\bibfield  {journal} {\bibinfo  {journal} {Phys.\ Rev.\ Lett.}\ }\textbf
  {\bibinfo {volume} {107}},\ \bibinfo {pages} {214101} (\bibinfo {year}
  {2011})}\BibitemShut {NoStop}%
\bibitem [{\citenamefont {Heller}(1991)}]{Heller:1991}%
  \BibitemOpen
  \bibfield  {author} {\bibinfo {author} {\bibfnamefont {E.~J.}\ \bibnamefont
  {Heller}},\ }\href {\doibase 10.1063/1.459848} {\bibfield  {journal}
  {\bibinfo  {journal} {J.~Chem.\ Phys.}\ }\textbf {\bibinfo {volume} {94}},\
  \bibinfo {pages} {2723} (\bibinfo {year} {1991})}\BibitemShut {NoStop}%
\bibitem [{\citenamefont {Zambrano}\ and\ \citenamefont {Ozorio~de
  Almeida}(2011)}]{Zambrano_Almeida:2011}%
  \BibitemOpen
  \bibfield  {author} {\bibinfo {author} {\bibfnamefont {E.}~\bibnamefont
  {Zambrano}}\ and\ \bibinfo {author} {\bibfnamefont {A.~M.}\ \bibnamefont
  {Ozorio~de Almeida}},\ }\href {\doibase 10.1103/PhysRevE.84.045201}
  {\bibfield  {journal} {\bibinfo  {journal} {Phys.\ Rev.~E}\ }\textbf
  {\bibinfo {volume} {84}},\ \bibinfo {pages} {045201(R)} (\bibinfo {year}
  {2011})}\BibitemShut {NoStop}%
\bibitem [{\citenamefont {Pollard}, \citenamefont {Lee},\ and\ \citenamefont
  {Mathies}(1990)}]{Pollard_Lee:1990}%
  \BibitemOpen
  \bibfield  {author} {\bibinfo {author} {\bibfnamefont {W.~T.}\ \bibnamefont
  {Pollard}}, \bibinfo {author} {\bibfnamefont {S.-Y.}\ \bibnamefont {Lee}}, \
  and\ \bibinfo {author} {\bibfnamefont {R.~A.}\ \bibnamefont {Mathies}},\
  }\href {\doibase 10.1063/1.457815} {\bibfield  {journal} {\bibinfo  {journal}
  {J.~Chem.\ Phys.}\ }\textbf {\bibinfo {volume} {92}},\ \bibinfo {pages}
  {4012} (\bibinfo {year} {1990})}\BibitemShut {NoStop}%
\bibitem [{\citenamefont {Stock}\ and\ \citenamefont
  {Domcke}(1992)}]{Stock_Domcke:1992}%
  \BibitemOpen
  \bibfield  {author} {\bibinfo {author} {\bibfnamefont {G.}~\bibnamefont
  {Stock}}\ and\ \bibinfo {author} {\bibfnamefont {W.}~\bibnamefont {Domcke}},\
  }\href {\doibase 10.1103/PhysRevA.45.3032} {\bibfield  {journal} {\bibinfo
  {journal} {Phys.\ Rev.~A}\ }\textbf {\bibinfo {volume} {45}},\ \bibinfo
  {pages} {3032} (\bibinfo {year} {1992})}\BibitemShut {NoStop}%
\bibitem [{\citenamefont {Tannor}(2004)}]{book_Tannor}%
  \BibitemOpen
  \bibfield  {author} {\bibinfo {author} {\bibfnamefont {D.~J.}\ \bibnamefont
  {Tannor}},\ }\href@noop {} {\emph {\bibinfo {title} {Introduction to Quantum
  Mechanics: A Time-Dependent Perspective}}}\ (\bibinfo  {publisher}
  {University Science Books},\ \bibinfo {year} {2004})\BibitemShut {NoStop}%
\bibitem [{\citenamefont {Gorin}\ \emph {et~al.}(2006)\citenamefont {Gorin},
  \citenamefont {Prosen}, \citenamefont {Seligman},\ and\ \citenamefont
  {\v{Z}nidari\v{c}}}]{Gorin_Prosen:2006}%
  \BibitemOpen
  \bibfield  {author} {\bibinfo {author} {\bibfnamefont {T.}~\bibnamefont
  {Gorin}}, \bibinfo {author} {\bibfnamefont {T.}~\bibnamefont {Prosen}},
  \bibinfo {author} {\bibfnamefont {T.~H.}\ \bibnamefont {Seligman}}, \ and\
  \bibinfo {author} {\bibfnamefont {M.}~\bibnamefont {\v{Z}nidari\v{c}}},\
  }\href {\doibase 10.1016/j.physrep.2006.09.003} {\bibfield  {journal}
  {\bibinfo  {journal} {Phys.\ Rep.}\ }\textbf {\bibinfo {volume} {435}},\
  \bibinfo {pages} {33} (\bibinfo {year} {2006})}\BibitemShut {NoStop}%
\bibitem [{\citenamefont {Jacquod}\ and\ \citenamefont
  {Petitjean}(2009)}]{Jacquod_Petitjean:2009}%
  \BibitemOpen
  \bibfield  {author} {\bibinfo {author} {\bibfnamefont {P.}~\bibnamefont
  {Jacquod}}\ and\ \bibinfo {author} {\bibfnamefont {C.}~\bibnamefont
  {Petitjean}},\ }\href {\doibase 10.1080/00018730902831009} {\bibfield
  {journal} {\bibinfo  {journal} {Adv.\ Phys.}\ }\textbf {\bibinfo {volume}
  {58}},\ \bibinfo {pages} {67} (\bibinfo {year} {2009})}\BibitemShut {NoStop}%
\bibitem [{\citenamefont {Pastawski}\ \emph {et~al.}(2000)\citenamefont
  {Pastawski}, \citenamefont {Levstein}, \citenamefont {Usaj}, \citenamefont
  {Raya},\ and\ \citenamefont {Hirschinger}}]{Pastawski_Levstein:2000}%
  \BibitemOpen
  \bibfield  {author} {\bibinfo {author} {\bibfnamefont {H.~M.}\ \bibnamefont
  {Pastawski}}, \bibinfo {author} {\bibfnamefont {P.~R.}\ \bibnamefont
  {Levstein}}, \bibinfo {author} {\bibfnamefont {G.}~\bibnamefont {Usaj}},
  \bibinfo {author} {\bibfnamefont {J.}~\bibnamefont {Raya}}, \ and\ \bibinfo
  {author} {\bibfnamefont {J.}~\bibnamefont {Hirschinger}},\ }\href {\doibase
  10.1016/S0378-4371(00)00146-1} {\bibfield  {journal} {\bibinfo  {journal}
  {Physica A}\ }\textbf {\bibinfo {volume} {283}},\ \bibinfo {pages} {166}
  (\bibinfo {year} {2000})}\BibitemShut {NoStop}%
\bibitem [{\citenamefont {Cucchietti}\ \emph {et~al.}(2003)\citenamefont
  {Cucchietti}, \citenamefont {Dalvit}, \citenamefont {Paz},\ and\
  \citenamefont {Zurek}}]{Cucchietti_Dalvit:2003}%
  \BibitemOpen
  \bibfield  {author} {\bibinfo {author} {\bibfnamefont {F.~M.}\ \bibnamefont
  {Cucchietti}}, \bibinfo {author} {\bibfnamefont {D.~A.~R.}\ \bibnamefont
  {Dalvit}}, \bibinfo {author} {\bibfnamefont {J.~P.}\ \bibnamefont {Paz}}, \
  and\ \bibinfo {author} {\bibfnamefont {W.~H.}\ \bibnamefont {Zurek}},\ }\href
  {\doibase 10.1103/PhysRevLett.91.210403} {\bibfield  {journal} {\bibinfo
  {journal} {Phys.\ Rev.\ Lett.}\ }\textbf {\bibinfo {volume} {91}},\ \bibinfo
  {pages} {210403} (\bibinfo {year} {2003})}\BibitemShut {NoStop}%
\bibitem [{\citenamefont {Gorin}, \citenamefont {Prosen},\ and\ \citenamefont
  {Seligman}(2004)}]{Gorin_Prosen:2004}%
  \BibitemOpen
  \bibfield  {author} {\bibinfo {author} {\bibfnamefont {T.}~\bibnamefont
  {Gorin}}, \bibinfo {author} {\bibfnamefont {T.}~\bibnamefont {Prosen}}, \
  and\ \bibinfo {author} {\bibfnamefont {T.~H.}\ \bibnamefont {Seligman}},\
  }\href {http://stacks.iop.org/1367-2630/6/i=1/a=020} {\bibfield  {journal}
  {\bibinfo  {journal} {New J.~Phys.}\ }\textbf {\bibinfo {volume} {6}},\
  \bibinfo {pages} {20} (\bibinfo {year} {2004})}\BibitemShut {NoStop}%
\bibitem [{\citenamefont {Petitjean}\ \emph {et~al.}(2007)\citenamefont
  {Petitjean}, \citenamefont {Bevilaqua}, \citenamefont {Heller},\ and\
  \citenamefont {Jacquod}}]{Petitjean_Bevilaqua:2007}%
  \BibitemOpen
  \bibfield  {author} {\bibinfo {author} {\bibfnamefont {C.}~\bibnamefont
  {Petitjean}}, \bibinfo {author} {\bibfnamefont {D.~V.}\ \bibnamefont
  {Bevilaqua}}, \bibinfo {author} {\bibfnamefont {E.~J.}\ \bibnamefont
  {Heller}}, \ and\ \bibinfo {author} {\bibfnamefont {P.}~\bibnamefont
  {Jacquod}},\ }\href {\doibase 10.1103/PhysRevLett.98.164101} {\bibfield
  {journal} {\bibinfo  {journal} {Phys.\ Rev.\ Lett.}\ }\textbf {\bibinfo
  {volume} {98}},\ \bibinfo {pages} {164101} (\bibinfo {year}
  {2007})}\BibitemShut {NoStop}%
\bibitem [{\citenamefont {Zimmermann}\ and\ \citenamefont
  {Van\'{\i}\v{c}ek}(2010)}]{Zimmermann_Vanicek:2010}%
  \BibitemOpen
  \bibfield  {author} {\bibinfo {author} {\bibfnamefont {T.}~\bibnamefont
  {Zimmermann}}\ and\ \bibinfo {author} {\bibfnamefont {J.}~\bibnamefont
  {Van\'{\i}\v{c}ek}},\ }\href {\doibase 10.1063/1.3451266} {\bibfield
  {journal} {\bibinfo  {journal} {J.~Chem.\ Phys.}\ }\textbf {\bibinfo {volume}
  {132}},\ \bibinfo {eid} {241101} (\bibinfo {year} {2010})}\BibitemShut
  {NoStop}%
\bibitem [{\citenamefont {Li}, \citenamefont {Mollica},\ and\ \citenamefont
  {Van\'{\i}\v{c}ek}(2009)}]{Li_Mollica:2009}%
  \BibitemOpen
  \bibfield  {author} {\bibinfo {author} {\bibfnamefont {B.}~\bibnamefont
  {Li}}, \bibinfo {author} {\bibfnamefont {C.}~\bibnamefont {Mollica}}, \ and\
  \bibinfo {author} {\bibfnamefont {J.}~\bibnamefont {Van\'{\i}\v{c}ek}},\
  }\href {\doibase 10.1063/1.3187240} {\bibfield  {journal} {\bibinfo
  {journal} {J.~Chem.\ Phys.}\ }\textbf {\bibinfo {volume} {131}},\ \bibinfo
  {eid} {041101} (\bibinfo {year} {2009})}\BibitemShut {NoStop}%
\bibitem [{\citenamefont {Zimmermann}\ \emph {et~al.}(2010)\citenamefont
  {Zimmermann}, \citenamefont {Ruppen}, \citenamefont {Li},\ and\ \citenamefont
  {Van\'{\i}\v{c}ek}}]{Zimmermann_Ruppen:2010}%
  \BibitemOpen
  \bibfield  {author} {\bibinfo {author} {\bibfnamefont {T.}~\bibnamefont
  {Zimmermann}}, \bibinfo {author} {\bibfnamefont {J.}~\bibnamefont {Ruppen}},
  \bibinfo {author} {\bibfnamefont {B.}~\bibnamefont {Li}}, \ and\ \bibinfo
  {author} {\bibfnamefont {J.}~\bibnamefont {Van\'{\i}\v{c}ek}},\ }\href
  {\doibase 10.1002/qua.22730} {\bibfield  {journal} {\bibinfo  {journal}
  {Int.~J.~Quant.\ Chem.}\ }\textbf {\bibinfo {volume} {110}},\ \bibinfo
  {pages} {2426} (\bibinfo {year} {2010})}\BibitemShut {NoStop}%
\bibitem [{\citenamefont {Berry}(1989)}]{Berry:1989}%
  \BibitemOpen
  \bibfield  {author} {\bibinfo {author} {\bibfnamefont {M.~V.}\ \bibnamefont
  {Berry}},\ }\href {\doibase 10.1098/rspa.1989.0052} {\bibfield  {journal}
  {\bibinfo  {journal} {Proc.\ Roy.\ Soc.\ London Sect.~A}\ }\textbf {\bibinfo
  {volume} {423}},\ \bibinfo {pages} {219} (\bibinfo {year}
  {1989})}\BibitemShut {NoStop}%
\bibitem [{\citenamefont {Ozorio~de Almeida}(1998)}]{Almeida:1998}%
  \BibitemOpen
  \bibfield  {author} {\bibinfo {author} {\bibfnamefont {A.~M.}\ \bibnamefont
  {Ozorio~de Almeida}},\ }\href {\doibase 10.1016/S0370-1573(97)00070-7}
  {\bibfield  {journal} {\bibinfo  {journal} {Phys.\ Rep.}\ }\textbf {\bibinfo
  {volume} {295}},\ \bibinfo {pages} {265} (\bibinfo {year}
  {1998})}\BibitemShut {NoStop}%
\bibitem [{\citenamefont {Bohigas}\ \emph {et~al.}(1995)\citenamefont
  {Bohigas}, \citenamefont {Giannoni}, \citenamefont {de~Almeida},\ and\
  \citenamefont {Schmit}}]{Bohigas_Giannoni:1995}%
  \BibitemOpen
  \bibfield  {author} {\bibinfo {author} {\bibfnamefont {O.}~\bibnamefont
  {Bohigas}}, \bibinfo {author} {\bibfnamefont {M.-J.}\ \bibnamefont
  {Giannoni}}, \bibinfo {author} {\bibfnamefont {A.~M.~O.}\ \bibnamefont
  {de~Almeida}}, \ and\ \bibinfo {author} {\bibfnamefont {C.}~\bibnamefont
  {Schmit}},\ }\href {\doibase 10.1088/0951-7715/8/2/005} {\bibfield  {journal}
  {\bibinfo  {journal} {Nonlinearity}\ }\textbf {\bibinfo {volume} {8}},\
  \bibinfo {pages} {203} (\bibinfo {year} {1995})}\BibitemShut {NoStop}%
\bibitem [{\citenamefont {Walton}\ and\ \citenamefont
  {Manolopoulos}(1996)}]{Walton_Manolopoulos:1996}%
  \BibitemOpen
  \bibfield  {author} {\bibinfo {author} {\bibfnamefont {A.~R.}\ \bibnamefont
  {Walton}}\ and\ \bibinfo {author} {\bibfnamefont {D.~E.}\ \bibnamefont
  {Manolopoulos}},\ }\href {\doibase 10.1080/00268979600100651} {\bibfield
  {journal} {\bibinfo  {journal} {Mol.\ Phys.}\ }\textbf {\bibinfo {volume}
  {87}},\ \bibinfo {pages} {961} (\bibinfo {year} {1996})}\BibitemShut
  {NoStop}%
\bibitem [{\citenamefont {Filinov}(1986)}]{Filinov:1986}%
  \BibitemOpen
  \bibfield  {author} {\bibinfo {author} {\bibfnamefont {V.~S.}\ \bibnamefont
  {Filinov}},\ }\href {\doibase 10.1016/S0550-3213(86)80034-7} {\bibfield
  {journal} {\bibinfo  {journal} {Nucl.\ Phys.~B}\ }\textbf {\bibinfo {volume}
  {271}},\ \bibinfo {pages} {717} (\bibinfo {year} {1986})}\BibitemShut
  {NoStop}%
\bibitem [{\citenamefont {Makri}\ and\ \citenamefont
  {Miller}(1987)}]{Makri_Miller:1987}%
  \BibitemOpen
  \bibfield  {author} {\bibinfo {author} {\bibfnamefont {N.}~\bibnamefont
  {Makri}}\ and\ \bibinfo {author} {\bibfnamefont {W.~H.}\ \bibnamefont
  {Miller}},\ }\href {\doibase 10.1016/0009-2614(87)80142-2} {\bibfield
  {journal} {\bibinfo  {journal} {Chem.\ Phys.\ Lett.}\ }\textbf {\bibinfo
  {volume} {139}},\ \bibinfo {pages} {10} (\bibinfo {year} {1987})}\BibitemShut
  {NoStop}%
\bibitem [{\citenamefont {Wang}, \citenamefont {Manolopoulos},\ and\
  \citenamefont {Miller}(2001)}]{Wang:2001}%
  \BibitemOpen
  \bibfield  {author} {\bibinfo {author} {\bibfnamefont {H.}~\bibnamefont
  {Wang}}, \bibinfo {author} {\bibfnamefont {D.~E.}\ \bibnamefont
  {Manolopoulos}}, \ and\ \bibinfo {author} {\bibfnamefont {W.~H.}\
  \bibnamefont {Miller}},\ }\href {\doibase 10.1063/1.1402992} {\bibfield
  {journal} {\bibinfo  {journal} {J.~Chem.\ Phys.}\ }\textbf {\bibinfo {volume}
  {115}},\ \bibinfo {pages} {6317} (\bibinfo {year} {2001})}\BibitemShut
  {NoStop}%
\bibitem [{\citenamefont {Widder}(1954)}]{Widder:1954}%
  \BibitemOpen
  \bibfield  {author} {\bibinfo {author} {\bibfnamefont {D.~V.}\ \bibnamefont
  {Widder}},\ }\href@noop {} {\bibfield  {journal} {\bibinfo  {journal} {Bull.\
  Amer.\ Math.\ Soc.}\ }\textbf {\bibinfo {volume} {60}},\ \bibinfo {pages}
  {444} (\bibinfo {year} {1954})}\BibitemShut {NoStop}%
\bibitem [{\citenamefont {Doll}, \citenamefont {Freeman},\ and\ \citenamefont
  {Gillan}(1988)}]{Doll:1988}%
  \BibitemOpen
  \bibfield  {author} {\bibinfo {author} {\bibfnamefont {J.}~\bibnamefont
  {Doll}}, \bibinfo {author} {\bibfnamefont {D.}~\bibnamefont {Freeman}}, \
  and\ \bibinfo {author} {\bibfnamefont {M.}~\bibnamefont {Gillan}},\ }\href
  {\doibase 10.1016/0009-2614(88)87380-9} {\bibfield  {journal} {\bibinfo
  {journal} {Chem.\ Phys.\ Lett.}\ }\textbf {\bibinfo {volume} {143}},\
  \bibinfo {pages} {277} (\bibinfo {year} {1988})}\BibitemShut {NoStop}%
\bibitem [{\citenamefont {Press}\ \emph {et~al.}(2007)\citenamefont {Press},
  \citenamefont {Teukolsky}, \citenamefont {Vetterling},\ and\ \citenamefont
  {Flannery}}]{Press_NumericalRecipes:2010}%
  \BibitemOpen
  \bibfield  {author} {\bibinfo {author} {\bibfnamefont {W.~H.}\ \bibnamefont
  {Press}}, \bibinfo {author} {\bibfnamefont {S.~A.}\ \bibnamefont
  {Teukolsky}}, \bibinfo {author} {\bibfnamefont {W.~T.}\ \bibnamefont
  {Vetterling}}, \ and\ \bibinfo {author} {\bibfnamefont {B.~P.}\ \bibnamefont
  {Flannery}},\ }\href
  {http://www.cambridge.org/gb/knowledge/isbn/item1174754/?site_locale=en_GB}
  {\emph {\bibinfo {title} {Numerical Recipes}}},\ \bibinfo {edition} {3rd}\
  ed.\ (\bibinfo  {publisher} {Cambridge University Press},\ \bibinfo {year}
  {2007})\BibitemShut {NoStop}%
\bibitem [{\citenamefont {Nocedal}\ and\ \citenamefont
  {Wright}(2006)}]{book_Nocedal}%
  \BibitemOpen
  \bibfield  {author} {\bibinfo {author} {\bibfnamefont {J.}~\bibnamefont
  {Nocedal}}\ and\ \bibinfo {author} {\bibfnamefont {S.~J.}\ \bibnamefont
  {Wright}},\ }\href@noop {} {\emph {\bibinfo {title} {Numerical
  optimization}}},\ \bibinfo {edition} {2nd}\ ed.\ (\bibinfo  {publisher}
  {Springer},\ \bibinfo {year} {2006})\BibitemShut {NoStop}%
\bibitem [{\citenamefont {Sun}\ and\ \citenamefont
  {Miller}(1999)}]{Sun_Miller:1999}%
  \BibitemOpen
  \bibfield  {author} {\bibinfo {author} {\bibfnamefont {X.}~\bibnamefont
  {Sun}}\ and\ \bibinfo {author} {\bibfnamefont {W.~H.}\ \bibnamefont
  {Miller}},\ }\href {\doibase 10.1063/1.478571} {\bibfield  {journal}
  {\bibinfo  {journal} {J.~Chem.\ Phys.}\ }\textbf {\bibinfo {volume} {110}},\
  \bibinfo {pages} {6635} (\bibinfo {year} {1999})}\BibitemShut {NoStop}%
\bibitem [{\citenamefont {Stock}\ \emph {et~al.}(1995)\citenamefont {Stock},
  \citenamefont {Woywod}, \citenamefont {Domcke}, \citenamefont {Swinney},\
  and\ \citenamefont {Hudson}}]{Stock_Woywod:1995}%
  \BibitemOpen
  \bibfield  {author} {\bibinfo {author} {\bibfnamefont {G.}~\bibnamefont
  {Stock}}, \bibinfo {author} {\bibfnamefont {C.}~\bibnamefont {Woywod}},
  \bibinfo {author} {\bibfnamefont {W.}~\bibnamefont {Domcke}}, \bibinfo
  {author} {\bibfnamefont {T.}~\bibnamefont {Swinney}}, \ and\ \bibinfo
  {author} {\bibfnamefont {B.~S.}\ \bibnamefont {Hudson}},\ }\href {\doibase
  10.1063/1.470689} {\bibfield  {journal} {\bibinfo  {journal} {J.~Chem.\
  Phys.}\ }\textbf {\bibinfo {volume} {103}},\ \bibinfo {pages} {6851}
  (\bibinfo {year} {1995})}\BibitemShut {NoStop}%
\bibitem [{\citenamefont {Duschinsky}(1937)}]{Duschinsky:1937}%
  \BibitemOpen
  \bibfield  {author} {\bibinfo {author} {\bibfnamefont {F.}~\bibnamefont
  {Duschinsky}},\ }\href@noop {} {\bibfield  {journal} {\bibinfo  {journal}
  {Acta Physicochim.\ URSS}\ }\textbf {\bibinfo {volume} {7}},\ \bibinfo
  {pages} {551} (\bibinfo {year} {1937})}\BibitemShut {NoStop}%
\bibitem [{\citenamefont {\"{O}zkan}(1990)}]{Ozkan:1990}%
  \BibitemOpen
  \bibfield  {author} {\bibinfo {author} {\bibfnamefont {{\.{I}}.}~\bibnamefont
  {\"{O}zkan}},\ }\href {\doibase 10.1016/0022-2852(90)90247-N} {\bibfield
  {journal} {\bibinfo  {journal} {J.~Mol.\ Spec.}\ }\textbf {\bibinfo {volume}
  {139}},\ \bibinfo {pages} {147} (\bibinfo {year} {1990})}\BibitemShut
  {NoStop}%
\bibitem [{\citenamefont {Meyer}, \citenamefont {Gatti},\ and\ \citenamefont
  {Worth}(2009)}]{book_MCTDH}%
  \BibitemOpen
  \bibinfo {editor} {\bibfnamefont {H.-D.}\ \bibnamefont {Meyer}}, \bibinfo
  {editor} {\bibfnamefont {F.}~\bibnamefont {Gatti}}, \ and\ \bibinfo {editor}
  {\bibfnamefont {G.~A.}\ \bibnamefont {Worth}},\ eds.,\ \href@noop {} {\emph
  {\bibinfo {title} {Multidimensional Quantum Dynamics: MCTDH Theory and
  Applications}}},\ \bibinfo {edition} {1st}\ ed.\ (\bibinfo  {publisher}
  {Wiley-VCH},\ \bibinfo {address} {Weinheim},\ \bibinfo {year}
  {2009})\BibitemShut {NoStop}%
\bibitem [{\citenamefont {Meyer}(1986)}]{Meyer:1986}%
  \BibitemOpen
  \bibfield  {author} {\bibinfo {author} {\bibfnamefont {H.-D.}\ \bibnamefont
  {Meyer}},\ }\href {\doibase 10.1063/1.450296} {\bibfield  {journal} {\bibinfo
   {journal} {J.~Chem.\ Phys.}\ }\textbf {\bibinfo {volume} {84}},\ \bibinfo
  {pages} {3147} (\bibinfo {year} {1986})}\BibitemShut {NoStop}%
\bibitem [{\citenamefont {Waterland}\ \emph {et~al.}(1988)\citenamefont
  {Waterland}, \citenamefont {Yuan}, \citenamefont {Martens}, \citenamefont
  {Gillilan},\ and\ \citenamefont {Reinhardt}}]{Waterland_Yuan:1988}%
  \BibitemOpen
  \bibfield  {author} {\bibinfo {author} {\bibfnamefont {R.~L.}\ \bibnamefont
  {Waterland}}, \bibinfo {author} {\bibfnamefont {J.-M.}\ \bibnamefont {Yuan}},
  \bibinfo {author} {\bibfnamefont {C.~C.}\ \bibnamefont {Martens}}, \bibinfo
  {author} {\bibfnamefont {R.~E.}\ \bibnamefont {Gillilan}}, \ and\ \bibinfo
  {author} {\bibfnamefont {W.~P.}\ \bibnamefont {Reinhardt}},\ }\href {\doibase
  10.1103/PhysRevLett.61.2733} {\bibfield  {journal} {\bibinfo  {journal}
  {Phys.\ Rev.\ Lett.}\ }\textbf {\bibinfo {volume} {61}},\ \bibinfo {pages}
  {2733} (\bibinfo {year} {1988})}\BibitemShut {NoStop}%
\bibitem [{\citenamefont {Eckhardt}, \citenamefont {Hose},\ and\ \citenamefont
  {Pollak}(1989)}]{Eckhardt_Hose:1989}%
  \BibitemOpen
  \bibfield  {author} {\bibinfo {author} {\bibfnamefont {B.}~\bibnamefont
  {Eckhardt}}, \bibinfo {author} {\bibfnamefont {G.}~\bibnamefont {Hose}}, \
  and\ \bibinfo {author} {\bibfnamefont {E.}~\bibnamefont {Pollak}},\ }\href
  {\doibase 10.1103/PhysRevA.39.3776} {\bibfield  {journal} {\bibinfo
  {journal} {Phys.\ Rev.~A}\ }\textbf {\bibinfo {volume} {39}},\ \bibinfo
  {pages} {3776} (\bibinfo {year} {1989})}\BibitemShut {NoStop}%
\bibitem [{\citenamefont {Bohigas}, \citenamefont {Tomsovic},\ and\
  \citenamefont {Ullmo}(1993)}]{Bohigas_Tomsovic:1993}%
  \BibitemOpen
  \bibfield  {author} {\bibinfo {author} {\bibfnamefont {O.}~\bibnamefont
  {Bohigas}}, \bibinfo {author} {\bibfnamefont {S.}~\bibnamefont {Tomsovic}}, \
  and\ \bibinfo {author} {\bibfnamefont {D.}~\bibnamefont {Ullmo}},\ }\href
  {\doibase 10.1016/0370-1573(93)90109-Q} {\bibfield  {journal} {\bibinfo
  {journal} {Phys.\ Rep.}\ }\textbf {\bibinfo {volume} {223}},\ \bibinfo
  {pages} {43} (\bibinfo {year} {1993})}\BibitemShut {NoStop}%
\bibitem [{\citenamefont {Revuelta}\ \emph {et~al.}(2012)\citenamefont
  {Revuelta}, \citenamefont {Vergini}, \citenamefont {Benito},\ and\
  \citenamefont {Borondo}}]{Revuelta_Vergini:2012}%
  \BibitemOpen
  \bibfield  {author} {\bibinfo {author} {\bibfnamefont {F.}~\bibnamefont
  {Revuelta}}, \bibinfo {author} {\bibfnamefont {E.~G.}\ \bibnamefont
  {Vergini}}, \bibinfo {author} {\bibfnamefont {R.~M.}\ \bibnamefont {Benito}},
  \ and\ \bibinfo {author} {\bibfnamefont {F.}~\bibnamefont {Borondo}},\ }\href
  {\doibase 10.1103/PhysRevE.85.026214} {\bibfield  {journal} {\bibinfo
  {journal} {Phys.\ Rev.~E}\ }\textbf {\bibinfo {volume} {85}},\ \bibinfo
  {pages} {026214} (\bibinfo {year} {2012})}\BibitemShut {NoStop}%
\bibitem [{\citenamefont {Li}\ \emph {et~al.}(1993)\citenamefont {Li},
  \citenamefont {Carter}, \citenamefont {Hirsch},\ and\ \citenamefont
  {Buenker}}]{Li_Carter:1993}%
  \BibitemOpen
  \bibfield  {author} {\bibinfo {author} {\bibfnamefont {Y.}~\bibnamefont
  {Li}}, \bibinfo {author} {\bibfnamefont {S.}~\bibnamefont {Carter}}, \bibinfo
  {author} {\bibfnamefont {G.}~\bibnamefont {Hirsch}}, \ and\ \bibinfo {author}
  {\bibfnamefont {R.~J.}\ \bibnamefont {Buenker}},\ }\href {\doibase
  10.1080/00268979300102131} {\bibfield  {journal} {\bibinfo  {journal} {Mol.\
  Phys.}\ }\textbf {\bibinfo {volume} {80}},\ \bibinfo {pages} {145} (\bibinfo
  {year} {1993})}\BibitemShut {NoStop}%
\bibitem [{\citenamefont {Baer}\ and\ \citenamefont
  {Kosloff}(1995)}]{Baer_Kosloff:1995}%
  \BibitemOpen
  \bibfield  {author} {\bibinfo {author} {\bibfnamefont {R.}~\bibnamefont
  {Baer}}\ and\ \bibinfo {author} {\bibfnamefont {R.}~\bibnamefont {Kosloff}},\
  }\href {\doibase 10.1021/j100009a011} {\bibfield  {journal} {\bibinfo
  {journal} {J.~Phys.\ Chem.}\ }\textbf {\bibinfo {volume} {99}},\ \bibinfo
  {pages} {2534} (\bibinfo {year} {1995})}\BibitemShut {NoStop}%
\bibitem [{\citenamefont {Brewer}, \citenamefont {Hulme},\ and\ \citenamefont
  {Manolopoulos}(1997)}]{Brewer_Hulme:1997}%
  \BibitemOpen
  \bibfield  {author} {\bibinfo {author} {\bibfnamefont {M.~L.}\ \bibnamefont
  {Brewer}}, \bibinfo {author} {\bibfnamefont {J.~S.}\ \bibnamefont {Hulme}}, \
  and\ \bibinfo {author} {\bibfnamefont {D.~E.}\ \bibnamefont {Manolopoulos}},\
  }\href {\doibase 10.1063/1.473532} {\bibfield  {journal} {\bibinfo  {journal}
  {J.~Chem.\ Phys.}\ }\textbf {\bibinfo {volume} {106}},\ \bibinfo {pages}
  {4832} (\bibinfo {year} {1997})}\BibitemShut {NoStop}%
\bibitem [{\citenamefont {Trotter}(1959)}]{Trotter:1959}%
  \BibitemOpen
  \bibfield  {author} {\bibinfo {author} {\bibfnamefont {H.~F.}\ \bibnamefont
  {Trotter}},\ }\href {\doibase 10.1090/S0002-9939-1959-0108732-6} {\bibfield
  {journal} {\bibinfo  {journal} {Proc.\ Amer.\ Math.\ Soc.}\ }\textbf
  {\bibinfo {volume} {10}},\ \bibinfo {pages} {545} (\bibinfo {year}
  {1959})}\BibitemShut {NoStop}%
\end{thebibliography}

\end{document}